\documentclass[useAMS,usenatbib,usegraphicx]{mn2e}
\usepackage{times,epsfig}
\title[MILES -- II. Stellar atmospheric
parameters]{Medium-resolution Isaac Newton Telescope library of empirical spectra -- II. The stellar atmospheric
parameters}

\author[Cenarro et
  al.]{A. J. Cenarro$^{1}$\thanks{E-mail:cen@astrax.fis.ucm.es}, 
   R. F. Peletier$^{2,3}$, 
   P. S\'anchez--Bl\'azquez$^{4}$, 
   S. O. Selam$^{5}$,
   E. Toloba$^{1}$,
\newauthor
   N. Cardiel$^{1}$, 
   J. Falc\'on--Barroso$^{6,7}$,
   J. Gorgas$^{1}$, 
   J. Jim\'enez-Vicente$^{8}$, 
   A. Vazdekis$^{9}$ \\
$^{1}$Departamento de F\'{\i}sica de la Tierra, Astronom\'{\i}a y Astrof\'{\i}sica II, Universidad Complutense de Madrid, 28040 Madrid, Spain\\
$^{2}$Kapteyn Astronomical Institute, University of Groningen, Postbus 800, 9700 AV Groningen, The Netherlands\\
$^{3}$School of Physics and Astronomy, University of Nottingham, University Park, Nottingham NG7 2RD, United Kingdom\\
$^{4}$Laboratoire d'Astrophysique, Ecole Polytechnique F\'ed\'erale de Lausanne (EPFL), Observatoire, CH-1290, Sauverny, Switzerland\\
$^{5}$Department of Astronomy and Space Siences, Faculty of Sciences, Ankara University, 06100 Tandogan/Ankara, Turkey \\
$^{6}$Sterrewacht Leiden, Niels Bohrweg 2, 2333 CA, Leiden, The Netherlands\\
$^{7}$European Space and Technology Centre (ESTEC), Keplerlaan 1, Postbus 299, 2200 AG Noordwijk, The Netherlands \\
$^{8}$Departamento de F\'\i sica Te\'orica y del Cosmos, Universidad de Granada, Avenida Fuentenueva s/n, 18071 Granada, Spain\\
$^{9}$Instituto de Astrof\'{\i}sica de Canarias, V\'{\i}a L\'actea s/n, E-38200, La Laguna, Tenerife, Spain}

\begin{document}

\date{Accepted 2006 October 12. Received 2006 October 11; in original form 2006 July 28}

\pagerange{\pageref{firstpage}--\pageref{lastpage}} \pubyear{2006}

\maketitle

\label{firstpage}

\begin{abstract}
We present a homogeneous set of stellar atmospheric parameters
($T_{\rm eff}$, $\log g$, [Fe/H]) for MILES, a new spectral stellar
library covering the range $\lambda\lambda 3525 - 7500$\,\AA\ at
2.3\,\AA\ (FWHM) spectral resolution. The library consists of 985
stars spanning a large range in atmospheric parameters, from super
metal-rich, cool stars to hot, metal-poor stars. The spectral
resolution, spectral type coverage and number of stars represent a
substantial improvement over previous libraries used in population
synthesis models. The atmospheric parameters that we present here are
the result of a previous, extensive compilation from the
literature. In order to construct a homogeneous dataset of atmospheric
parameters we have taken the sample of stars of Soubiran, Katz \&
Cayrel, which has very well determined fundamental parameters, as the
standard reference system for our field stars, and have calibrated and
bootstrapped the data from other papers against it. The atmospheric
parameters for our cluster stars have also been revised and updated
according to recent metallicity scales, colour-temperature relations
and improved set of isochrones.
\end{abstract}

\begin{keywords}
atlases -- stars: fundamental parameters -- globular clusters: general
-- galaxies: stellar content
\end{keywords}

\section{Introduction}

This paper is the second one in a series whose ultimate goal is to
provide single stellar population (SSP) models in the optical spectral
range on the basis of MILES (S\'anchez-Bl\'azquez et al.~2006;
Paper~I). MILES is a medium-resolution (2.3\,\AA\ FWHM) spectral
stellar library in the region $\lambda\lambda 3500 - 7500$\,\AA\
consisting of 985 stars with an unprecedented coverage of stellar
atmospheric parameters. The present paper is dedicated to provide a
{\sl homogenized} set of effective temperatures $T_{\rm eff}$, surface
gravities $\log g$, and metallicities [Fe/H], for all the library
stars, parameters necessary to make reliable stellar population model
predictions. In the third paper of the series (Vazdekis et al.~2007,
in preparation; Paper~III), the new stellar population models at the
resolution and spectral range of MILES are going to be presented.

Stellar population models use stellar evolution theory, which predicts
the fundamental stellar atmospheric parameters (such as $T_{\rm eff}$,
$\log g$ and [Fe/H]) of the stars belonging to a stellar population of
given age and metallicity. The relative contribution of the stars in
the different evolutionary states is calculated by integrating along
the mass distribution, assuming a given initial mass function. To make
a high-quality stellar population model, one needs reliable stellar
interior models --that include as many phases in the life of a star as
possible-- as well as a trustworthy conversion between the stellar
parameters and the spectrum of the star.

In the past, people generally employed absorption line-strength
indices when trying to constrain the stellar populations of galaxies,
mostly using the Lick/IDS system (Gorgas et al.~1993, hereafter G93;
Worthey et al.~1994; hereafter WOR). For each star across the
isochrone, an index value was determined from its corresponding three
stellar atmospheric parameters (sometimes using colours instead of
$T_{\rm eff}$) using the so-called fitting functions, i.e. polynomial
functions relating the stellar parameters to the measured
pseudo-equivalent widths (G93; WOR; Worthey \& Ottaviani 1997). At
present, a new generation of stellar population models goes beyond the
prediction of individual features as these models synthesize at once a
full spectral energy distribution (SED; e.g. Vazdekis 1999; Vazdekis
et al.~2003, hereafter VAZ03; Bruzual \& Charlot 2003). For every set
of parameters, they take the corresponding stellar spectrum from an
observational or theoretical stellar library.

The price for predicting SEDs and keeping full information along the
spectrum is the necessity of ensuring a much higher quality for the
input ingredients for these models: the atmospheric parameters
coverage, the spectral resolution, and the relative flux calibration
of the library stars, are essential issues to be taken into
account. This is why we were motivated to perform the MILES project
and obtain a new set of observed input spectra of stars of all
spectral types, luminosity classes and metallicities at a spectral
resolution that is high enough to ensure that the population synthesis
of most galaxies --except for very low mass dwarfs-- is not limited by
the models but by the intrinsic broadening of the data. Although there
exist other comparable stellar libraries available in the literature
(see Paper~I for an in-depth comparison), the better parameter
coverage, the accurate flux calibration, as well as the large spectral
range of MILES provide a quite significant improvement.

As regards to the atmospheric parameters, while not so critical for
those models predictions based on fitting functions --in which
interpolations within the parameter space are immediate--, stellar
libraries covering the atmospheric parameters in an ample and
homogeneous way are essential to synthetise reliable, integrated
spectra over a wide range of ages an metallicities. In this sense, a
previous, thorough selection of stars was carried out for this project
(see Paper~I). Moreover, it is clear that uncertainties in the input
atmospheric parameters of the library stars have important
implications on the reliability and accuracy of the model predictions
(e.g. Gorgas et al.~1999). In the literature, atmospheric parameters
for most previous library stars are either taken from the most recent
bibliographic sources at that time or assigned as average values of
the existing determinations, without checking in any case whether they
are on a completely homogeneous system. A common practice is to use
straight means from previous parameter compilations (e.g. Cayrel de
Strobel et al.~1997), even though the individual analyses do not
necessarily all have the same quality or are mutually independent. We
refer the reader to the work of Soubiran, Katz \& Cayrel (1998;
hereafter SKC) for a thorough discussion of these and related
problems. Furthermore, systematic deviations among different
bibliographic sources may exist due to the different approaches for
measuring atmospheric parameters.

Because of the above limitations, a special effort was started in
Cenarro et al.~(2001b; herafter CEN01b) in order to construct a
homogeneous atmospheric parameter system for the stars in the near-IR,
CaT stellar library (Cenarro et al.~2001a; hereafter CEN01a). The
present paper may be considered as a step forward in the work and
procedure established in CEN01b. In this sense, we have derived an
enlarged, homogeneous set of stellar atmospheric parameters for most
stars in MILES. Section~2 presents the working procedure carried out
in this paper to determine the atmospheric parameters of field stars,
including data compilation and the calibration of the different
bibliographic sources with respect to a standard, reference system.
In Section~3, we have also recomputed the atmospheric parameters of
all the cluster stars in MILES making use of colour-temperature
relations and apropriate isochrones for each individual cluster, in a
similar way as in G93. A brief summary of the paper is presented in
Section\,4 and, finally, tables containing the newly derived
atmospheric parameters for the MILES stars are given in Appendix~A.

\section{Atmospheric parameters for field stars}

Following the procedure carried out in CEN01b to determine a
homogeneous set of atmospheric parameters for the field stars in the
near-IR stellar library (CEN01a), one of the main goals of this paper
is to construct a larger, homogenous set of atmospheric parameters for
such stars in MILES. In the following paragraphs in this section, we
explain the method used in this paper. For a more detailed explanation
of the working procedure, we refer the reader to CEN01b. In short, it
can be itemized in the following steps: i) selection of a
high-quality, standard reference of atmospheric parameters, ii)
bibliographic compilation of atmospheric parameters for the library
stars, iii) calibration and correction of systematic differences
between the different sources and the standard, reference system, and
iv) determination of averaged, final atmospheric parameters for the
library stars from all those references corrected onto the reference
system.

\subsection{The reference system}

In order to establish a homogeneous system of atmospheric parameters,
it is necessary to define an appropriate, initial reference system
against which other sources are calibrated and corrected for
systematic diferences.  Bearing in mind that the final purpose of this
series of papers is stellar population modeling of SEDs, we are
basically interested in ensuring that stars with very similar spectra
have the same atmospheric parameters and the other way round. This is
why, as in CEN01b, we have selected the work by SKC as our initial
standard source, since it computes self-consistent atmospheric
parameters for a total of 211 echelle spectra of stars with 4000~K
$<T_{\rm eff}<$ 6300~K over a wide range of $\log g$ and [Fe/H]. We
refer the reader to Katz et al.~(1998) for a detailed explanation of
the spectroscopic method followed in SKC to derive atmospheric
parameters.

\subsection{Parameter compilation from bibliographic sources}

Given that the stellar sample in SKC does not comprise all stars in
MILES, we updated the previous, extensive compilation of atmospheric
parameters in CEN01b by including data from recent publications and
extending the search to all field stars in MILES. Overall, it attains
20295 records, even though not all of them were finally employed to
derive the final parameters.  The catalogue of [Fe/H] determinations
of F, G and K-type stars of Cayrel de Strobel et al.~(2001) --that
contains parameters for more than 3000 stars from 378 different
sources up to 2000-- was our starting point. The compilation was
enlarged with several additional sources to account for the most
recent determinations as well as to include atmospheric parameters for
early and very late spectral types which are not included in the above
catalogue. It must be noted that, even for stars with data in Cayrel
de Strobel et al.~(2001), we checked the original data sources to
exclude references that simply quoted previous determinations.

\subsection{Calibration and correction of bibliographic sources}

\begin{scriptsize}                                                                                       
\begin{table}
\centering{ 
\caption{Calibrations of bibliographic sources to convert their
effective temperatures onto the reference system. Column description:
Code (reference code of the bibliographic source; see
Table~\ref{calrefer}); M (method used to derive temperatures in that
source: (a) Infrared flux method; (b) spectroscopic methods; (c) from
colour relations); N (number of stars in common with the standard
source); Fit (type of calibration applied to correct the original
data; s: straight line; o: offset; n: none); S (standard source; 1:
SKC; 2: SKC \& RF1); $\sigma$ (r.m.s. standard deviation from the
fit); $A$ and $B$ (independent term and slope of the applied fit; $p =
A + B\,p_{\rm ref}$, with $p$ and $p_{\rm ref}$ being generic
atmospheric parameters --$T_{\rm eff}$ in this case-- from the source
and the reference system respectively); $T_{\rm eff}$ (temperature
validity regime of the fit). Values from JON and WOR only include
original determinations, that is, parameters taken from other sources
were not employed (this also holds for Tables~2 and~3).}
\label{caltemp}
\begin{tabular}{@{}l@{}c@{}r@{}c@{}crrlc@{}}
\hline
 Code & M & N  &\ \ \ Fit&\ \ \ \ S& {\large $\sigma$}\ \ &$A$\,\ &\ $B$& {\it T}$_{\rm eff}$ \\
\hline                
AAM   &\ a   \ \ \ &  67 &\ \ \ \ n\ \ &\ \ \ \ 1&  98.0 &    0.0  & 1.0    &4300 , 6400\\                        
AFG   &\ b   \ \ \ &  30 &\ \ \ \ n\ \ &\ \ \ \ 1& 124.0 &    0.0  & 1.0    &5600 , 6400\\                        
BAL   &\ c   \ \ \ &  24 &\ \ \ \ o\ \ &\ \ \ \ 2& 107.2 &   47.5  & 1.0    &5900 , 6400\\
BLG   &\ a   \ \ \ &  27 &\ \ \ \ o\ \ &\ \ \ \ 2&  61.8 &   51.0  & 1.0    &4200 , 6250\\
BOV   &\ c   \ \ \ &  18 &\ \ \ \ n\ \ &\ \ \ \ 2& 132.1 & --74.0  & 1.0    &4150 , 5450\\
BSL   &\ c   \ \ \ &  39 &\ \ \ \ s\ \ &\ \ \ \ 1&  66.0 &  396.5  & 0.9118 &4000 , 5100\\                        
CLG   &\ c   \ \ \ &  37 &\ \ \ \ o\ \ &\ \ \ \ 2&  54.0 &  126.0  & 1.0    &4600 , 6100\\
CLL   &\ c   \ \ \ &  40 &\ \ \ \ n\ \ &\ \ \ \ 1&  76.0 &    0.0  & 1.0    &4600 , 6300\\                        
CNZ   &\ c   \ \ \ &  30 &\ \ \ \ o\ \ &\ \ \ \ 2&  81.8 & --41.5  & 1.0    &5600 , 6300\\
CTL   &\ bc  \ \ \ &  17 &\ \ \ \ s\ \ &\ \ \ \ 2&  48.0 &  928.5  & 0.8347 &5200 , 6350\\   
CTS   &\ b   \ \ \ &  18 &\ \ \ \ o\ \ &\ \ \ \ 2&  90.5 &--111.5  & 1.0    &4500 , 5050\\   
EAG   &\ c   \ \ \ &  36 &\ \ \ \ o\ \ &\ \ \ \ 1&  60.0 &   39.9  & 1.0    &5650 , 6350\\                        
FHR   &\ b   \ \ \ &  35 &\ \ \ \ o\ \ &\ \ \ \ 2&  81.6 &   53.7  & 1.0    &5000 , 6350\\
FLB   &\ b   \ \ \ &  61 &\ \ \ \ n\ \ &\ \ \ \ 1& 114.0 &    0.0  & 1.0    &4200 , 6400\\
FRC   &\ c   \ \ \ &  15 &\ \ \ \ o\ \ &\ \ \ \ 2&  65.0 & --61.0  & 1.0    &3750 , 5150\\
FUH   &\ -   \ \ \ &  25 &\ \ \ \ o\ \ &\ \ \ \ 2&  63.3 &   45.0  & 1.0    &5200 , 6400\\
GCC   &\ c   \ \ \ &  65 &\ \ \ \ s\ \ &\ \ \ \ 1&  86.0 &--178.8  & 1.0397 &4100 , 6500\\                        
GGR   &\ b   \ \ \ &  16 &\ \ \ \ o\ \ &\ \ \ \ 2&  37.0 &   64.0  & 1.0    &4300 , 5100\\
GRJ   &\ b   \ \ \ &  35 &\ \ \ \ o\ \ &\ \ \ \ 2& 108.9 &   50.0  & 1.0    &4300 , 6300\\
GRO   &\ c   \ \ \ &  15 &\ \ \ \ o\ \ &\ \ \ \ 2& 126.3 &  110.0  & 1.0    &3950 , 5600\\
GRS   &\ c   \ \ \ &  25 &\ \ \ \ n\ \ &\ \ \ \ 1& 116.0 &    0.0  & 1.0    &3800 , 6100\\                        
GRT   &\ c   \ \ \ &  17 &\ \ \ \ s\ \ &\ \ \ \ 2& 101.8 &  424.1  & 0.9178 &4050 , 6250\\
GSC   &\ c   \ \ \ &  34 &\ \ \ \ o\ \ &\ \ \ \ 2&  92.5 &  129.0  & 1.0    &4200 , 6150\\
HEA   &\ c   \ \ \ &  26 &\ \ \ \ s\ \ &\ \ \ \ 2&  69.1 &  887.1  & 0.8    &5000 , 6300\\
HWA   &\ c   \ \ \ &  17 &\ \ \ \ o\ \ &\ \ \ \ 2& 105.0 &--663.6  & 1.0    &3750 , 5200\\
JON   &\ c   \ \ \ & 105 &\ \ \ \ n\ \ &\ \ \ \ 2&  87.5 &    0.0  & 1.0    &4200 , 5300\\
KSP   &\ b   \ \ \ &  17 &\ \ \ \ o\ \ &\ \ \ \ 2&  81.3 &   64.0  & 1.0    &4300 , 6050\\
LAI   &\ c   \ \ \ &  73 &\ \ \ \ o\ \ &\ \ \ \ 2&  76.7 & --37.0  & 1.0    &4750 , 6300\\
LBO   &\ c   \ \ \ &  21 &\ \ \ \ n\ \ &\ \ \ \ 2& 195.3 &    0.0  & 1.0    &4250 , 6100\\
LCH   &\ c   \ \ \ &  35 &\ \ \ \ o\ \ &\ \ \ \ 2&  61.7 & --63.0  & 1.0    &3850 , 5050\\
LRS   &\ b   \ \ \ &  19 &\ \ \ \ o\ \ &\ \ \ \ 2&  85.0 &  242.0  & 1.0    &3900 , 5350\\
MAS   &\ c   \ \ \ &  38 &\ \ \ \ s\ \ &\ \ \ \ 1&  83.0 & 2852.0  & 0.5450 &5900 , 6300\\
MCW   &\ c   \ \ \ &  62 &\ \ \ \ n\ \ &\ \ \ \ 1&  86.0 &    0.0  & 1.0    &3900 , 5900\\
MEH   &\ b   \ \ \ &  22 &\ \ \ \ o\ \ &\ \ \ \ 2& 167.6 &   21.5  & 1.0    &3900 , 6100\\
MGN   &\ c   \ \ \ &  18 &\ \ \ \ n\ \ &\ \ \ \ 2&  65.0 &    0.0  & 1.0    &5600 , 6300\\
NHS   &\ c   \ \ \ &  33 &\ \ \ \ n\ \ &\ \ \ \ 2&  92.2 &    0.0  & 1.0    &4700 , 6350\\
OIN   &\ b   \ \ \ &  18 &\ \ \ \ n\ \ &\ \ \ \ 2& 118.2 &    0.0  & 1.0    &3600 , 5400\\
ONS   &\ b   \ \ \ &  16 &\ \ \ \ o\ \ &\ \ \ \ 2&  58.15&   41.0  & 1.0    &4000 , 5650\\
PET   &\ c   \ \ \ &  29 &\ \ \ \ o\ \ &\ \ \ \ 2& 110.6 & --75.0  & 1.0    &4450 , 6400\\
PSB   &\ c   \ \ \ &  26 &\ \ \ \ s\ \ &\ \ \ \ 1& 101.0 &  517.7  & 0.9042 &4300 , 6000\\
PSK   &\ bc  \ \ \ &  32 &\ \ \ \ n\ \ &\ \ \ \ 2& 113.3 &    0.0  & 1.0    &4050 , 5550\\
RBM   &\ bc  \ \ \ &  32 &\ \ \ \ o\ \ &\ \ \ \ 2&  87.0 & --33.5  & 1.0    &5200 , 6200\\
SAH   &\ a   \ \ \ &  16 &\ \ \ \ n\ \ &\ \ \ \ 2&  30.0 &    0.0  & 1.0    &5800 , 6400\\
SIC   &\ b   \ \ \ &  22 &\ \ \ \ n\ \ &\ \ \ \ 2& 140.4 &    0.0  & 1.0    &4150 , 6350\\
TAY   &\ abc \ \ \ &  62 &\ \ \ \ s\ \ &\ \ \ \ 1&  92.0 & 1075.9  & 0.8166 &4800 , 6200\\
TID   &\ b   \ \ \ &  35 &\ \ \ \ n\ \ &\ \ \ \ 1&  75.0 &    0.0  & 1.0    &4300 , 6300\\
TLA   &\ b   \ \ \ &  27 &\ \ \ \ o\ \ &\ \ \ \ 2& 110.5 &   66.0  & 1.0    &4200 , 5400\\
TLL   &\ c   \ \ \ &  31 &\ \ \ \ o\ \ &\ \ \ \ 2&  99.2 & --40.0  & 1.0    &4700 , 6250\\
WOR   &\ c   \ \ \ &  44 &\ \ \ \ n\ \ &\ \ \ \ 1&  74.0 &    0.0  & 1.0    &4100 , 6100\\
\hline                                          
\end{tabular}                                                                                      
}
\end{table}
\end{scriptsize}

\begin{scriptsize}                                                                                       
\begin{table}                                                                                            
\centering{ 
\caption{Calibrations of bibliographic sources to convert their surface gravities
onto the reference system. Columns are the same as in
Table~\ref{caltemp}. Methods employed to derive gravities: (a)
spectroscopic method, (b) physical method (parallaxes), (c) physical
method (luminosities from photometric indices), (d) physical method
(luminosities from Ca K line), (e) photometric, and (f) other.}
\label{calgrav}
\begin{tabular}{@{}lcrcccr@{}llc@{}}
\hline
 Code & M & N &Fit&S & {\large $\sigma$}&\multicolumn{2}{c}{$A$}&\ $B$& log\,{\it g}\\
\hline
AFG & a  &  30 &n&1  &  0.27   &    0.&0    &  1.0    &   2.5 ,  4.8    \\                                                
BAL & c  &  30 &o&2  &  0.13   &    0.&095  &  1.0    &   3.8 ,  4.3    \\   
BSL & cd &  39 &n&1  &  0.19   &    0.&0    &  1.0    &   1.4 ,  3.9    \\                                                
CLG & be &  33 &o&2  &  0.177  &    0.&16   &  1.0    &   2.3 ,  4.9    \\
CNZ & b  &  31 &s&2  &  0.08   &    1.&767  &  0.581  &   3.9 ,  4.5    \\   
CTL & bc &  18 &o&2  &  0.154  &  --0.&085  &  1.0    &   3.5 ,  4.7    \\   
CTS & a  &  18 &n&2  &  0.275  &    0.&0    &  1.0    &   1.7 ,  3.3    \\ 
EAG & f  &  36 &o&1  &  0.12   &    0.&042  &  1.0    &   3.9 ,  4.6    \\                                                
FHR & a  &  37 &n&2  &  0.12   &    0.&0    &  1.0    &   3.1 ,  4.7    \\ 
FLB & a  &  61 &n&1  &  0.25   &    0.&0    &  1.0    &   0.5 ,  4.9    \\
FRC & e  &  15 &s&2  &  0.256  &    0.&6465 &  0.6164 &   1.3 ,  2.9    \\
FUH & -  &  23 &s&2  &  0.11   &    0.&961  &  0.760  &   3.5 ,  4.7    \\
GCC & a  &  65 &s&1  &  0.24   &  --0.&200  &  1.077  &   0.0 ,  5.2    \\                                                
GGR & a  &  16 &o&2  &  0.305  &  --0.&365  &  1.0    &   0.9 ,  3.2    \\
GRS & b  &  24 &o&1  &  0.30   &    0.&139  &  1.0    &   0.7 ,  4.5    \\                                                
GRT & a  &  16 &s&2  &  0.315  &    0.&3381 &  0.8076 &   1.0 ,  4.4    \\
GSC & b  &  34 &o&2  &  0.20   &    0.&145  &  1.0    &   0.5 ,  5.0    \\
HEA & b  &  23 &n&2  &  0.17   &    0.&0    &  1.0    &   3.1 ,  4.7    \\
HWA & a  &  17 &o&2  &  0.24   &  --0.&23   &  1.0    &   1.7 ,  2.9    \\
KNK & e  &  28 &o&1  &  0.14   &    0.&075  &  1.0    &   4.0 ,  4.7    \\                                                
KSP & a  &  15 &o&2  &  0.281  &    0.&14   &  1.0    &   0.6 ,  4.5    \\
LAI & ab &  72 &n&2  &  0.25   &    0.&0    &  1.0    &   2.1 ,  5.1    \\
LBO & a  &  17 &n&2  &  0.45   &    0.&0    &  1.0    &   0.0 ,  3.9    \\
LCH & ad &  35 &o&2  &  0.39   &  --0.&420  &  1.0    &   0.2 ,  3.1    \\
LRS & a  &  18 &o&2  &  0.238  &    0.&33   &  1.0    &   1.0 ,  4.0    \\
MAM & b  &  16 &n&2  &  0.092  &    0.&0    &  1.0    &   3.5 ,  4.7    \\
MAS & e  &  38 &o&1  &  0.40   &    0.&247  &  1.0    &   3.8 ,  5.0    \\
MCW & bd &  62 &o&1  &  0.21   &    0.&233  &  1.0    &   1.6 ,  4.2    \\
MGN & a  &  18 &o&2  &  0.159  &  --0.&295  &  1.0    &   3.0 ,  4.4    \\
NHS & b  &  33 &o&2  &  0.17   &    0.&14   &  1.0    &   3.1 ,  4.8    \\
OIN & a  &  17 &s&2  &  0.179  &    0.&5467 &  0.8392 &   1.4 ,  4.7    \\
PSK & cf &  29 &n&2  &  0.26   &    0.&0    &  1.0    &   0.1 ,  3.0    \\
SAH & e  &  20 &s&2  &  0.065  &    1.&972  &  0.5449 &   3.8 ,  4.5    \\
TID & a  &  35 &o&1  &  0.13   &    0.&13   &  1.0    &   1.9 ,  4.8    \\
TLA & a  &  23 &n&2  &  0.29   &    0.&0    &  1.0    &   0.5 ,  4.8    \\
TLL & a  &  31 &n&2  &  0.26   &    0.&0    &  1.0    &   2.5 ,  5.1    \\
WOR & f  &  34 &n&1  &  0.33   &    0.&0    &  1.0    &   1.0 ,  4.8    \\
\hline                                          
\end{tabular}                                                                                      
}
\end{table}
\end{scriptsize}

\begin{scriptsize}                                                                                       
\begin{table}                                                                                            
\centering{ 
\caption{Calibrations of bibliographic sources to convert their metallicities 
onto the reference system. Columns are the same as in
Table~\ref{caltemp}. Methods employed to compute metallicities:
(a) high resolution ($< 0.5$\,\AA) spectroscopy, (b) mid resolution
($> 0.5$\,\AA) spectroscopy, (c) photometry, and (d) spectrophotometry.}
\label{calmetal}
\begin{tabular}{@{}l@{}c@{}r@{}c@{}ccr@{}llr@{}r@{}}
\hline
 Code & M & N  &\ \ \ Fit&\ \ \ \ \ S & {\large $\sigma$}& &$A$&\ \ $B$& \multicolumn{2}{c}{\ \ [Fe/H]}\\
\hline
AAM &\ \ ac\ \ \ &\  68&\ \ \ \ s\ \ &\ \ \ \ 1&  0.22  &--0.&006 & 1.065 &--3.0 ,&\  +0.4  \\  
AFG &\ \  a\ \ \ &\  30&\ \ \ \ s\ \ &\ \ \ \ 1&  0.13  &--0.&120 & 0.858 &--2.5 ,&\ --0.4  \\  
BAL &\ \  a\ \ \ &\  31&\ \ \ \ o\ \ &\ \ \ \ 2&  0.11  &--0.&060 & 1.0   &--0.7 ,&\  +0.4  \\  
BKP &\ \  b\ \ \ &\  27&\ \ \ \ s\ \ &\ \ \ \ 1&  0.21  &--0.&324 & 0.829 &--3.1 ,&\ --1.0  \\  
BSL &\ \  a\ \ \ &\  39&\ \ \ \ n\ \ &\ \ \ \ 1&  0.19  &  0.&0   & 1.0   &--0.8 ,&\  +0.5  \\  
CGC &\ \  a\ \ \ &\  36&\ \ \ \ o\ \ &\ \ \ \ 2&  0.13  &  0.&125 & 1.0   &--2.6 ,&\ --0.7  \\
CLG &\ \ ac\ \ \ &\  37&\ \ \ \ o\ \ &\ \ \ \ 2&  0.153 &  0.&19  & 1.0   &--2.2 ,&\ --0.1  \\
CLL &\ \  a\ \ \ &\  41&\ \ \ \ s\ \ &\ \ \ \ 1&  0.10  &  0.&029 & 1.070 &--2.7 ,&\  +0.2  \\  
CNZ &\ \  a\ \ \ &\  33&\ \ \ \ s\ \ &\ \ \ \ 2&  0.08  &--0.&096 & 0.798 &--1.2 ,&\  +0.2  \\
CTL &\ \  a\ \ \ &\  18&\ \ \ \ n\ \ &\ \ \ \ 2&  0.081 &  0.&0   & 1.0   &--0.9 ,&\  +0.4  \\
CTS &\ \  a\ \ \ &\  18&\ \ \ \ s\ \ &\ \ \ \ 2&  0.116 &--0.&2466& 0.5662&--1.0 ,&\  +0.2  \\
EAG &\ \  a\ \ \ &\  36&\ \ \ \ s\ \ &\ \ \ \ 1&  0.05  &--0.&047 & 0.925 &--1.1 ,&\  +0.2  \\  
FHR &\ \  a\ \ \ &\  35&\ \ \ \ n\ \ &\ \ \ \ 2&  0.09  &  0.&0   & 1.0   &--2.1 ,&\  +0.4  \\
FLB &\ \  a\ \ \ &\  61&\ \ \ \ o\ \ &\ \ \ \ 1&  0.14  &  0.&10  & 1.0   &--3.0 ,&\ --0.3  \\ 
FRA &\ \  a\ \ \ &\  19&\ \ \ \ o\ \ &\ \ \ \ 2&  0.119 &--0.&09  & 1.0   &--2.7 ,&\  +0.1  \\
FRC &\ \  a\ \ \ &\  15&\ \ \ \ n\ \ &\ \ \ \ 2&  0.183 &  0.&0   & 1.0   &--1.1 ,&\  +0.3  \\
FUH &\ \  -\ \ \ &\  23&\ \ \ \ n\ \ &\ \ \ \ 2&  0.08  &  0.&0   & 1.0   &--2.2 ,&\  +0.5  \\
GCC &\ \  a\ \ \ &\  65&\ \ \ \ s\ \ &\ \ \ \ 1&  0.10  &--0.&002 & 0.947 &--3.0 ,&\  +0.2  \\  
GGR &\ \  a\ \ \ &\  16&\ \ \ \ o\ \ &\ \ \ \ 2&  0.155 &  0.&105 & 1.0   &--0.6 ,&\  +0.3  \\
GRO &\ \  a\ \ \ &\  19&\ \ \ \ s\ \ &\ \ \ \ 2&  0.138 &--0.&1274& 0.8359&--2.8 ,&\ --0.1  \\
GRS &\ \  a\ \ \ &\  25&\ \ \ \ n\ \ &\ \ \ \ 1&  0.18  &  0.&0   & 1.0   &--2.4 ,&\  +0.2  \\  
GRT &\ \  a\ \ \ &\  17&\ \ \ \ o\ \ &\ \ \ \ 2&  0.117 &--0.&12  & 1.0   &--2.4 ,&\ --0.1  \\
GSC &\ \  a\ \ \ &\  34&\ \ \ \ o\ \ &\ \ \ \ 2&  0.09  &  0.&094 & 1.0   &--2.2 ,&\ --0.8  \\
HEA &\ \  a\ \ \ &\  23&\ \ \ \ n\ \ &\ \ \ \ 2&  0.18  &  0.&0   & 1.0   &--1.1 ,&\  +0.4  \\
HWA &\ \  a\ \ \ &\  17&\ \ \ \ o\ \ &\ \ \ \ 2&  0.137 &--0.&14  & 1.0   &--0.8 ,&\  +0.3  \\
JON &\ \  d\ \ \ &\  98&\ \ \ \ o\ \ &\ \ \ \ 2&  0.12  &  0.&075 & 1.0   &--1.0 ,&\  +0.6  \\
KNK &\ \  c\ \ \ &\  32&\ \ \ \ s\ \ &\ \ \ \ 1&  0.09  &--0.&036 & 0.911 &--2.1 ,&\  +0.2  \\  
KSP &\ \  a\ \ \ &\  17&\ \ \ \ n\ \ &\ \ \ \ 2&  0.14  &  0.&0   & 1.0   &--3.0 ,&\ --1.3  \\
LAI &\ \  b\ \ \ &\  72&\ \ \ \ o\ \ &\ \ \ \ 2&  0.16  &--0.&082 & 1.0   &--2.6 ,&\  +0.5  \\
LBO &\ \  a\ \ \ &\  27&\ \ \ \ o\ \ &\ \ \ \ 2&  0.12  &  0.&140 & 1.0   &--3.0 ,&\ --0.5  \\
LCH &\ \  a\ \ \ &\  33&\ \ \ \ s\ \ &\ \ \ \ 2&  0.16  &--0.&037 & 0.752 &--0.5 ,&\  +0.3  \\
LRS &\ \  a\ \ \ &\  19&\ \ \ \ n\ \ &\ \ \ \ 2&  0.109 &  0.&0   & 1.0   &--0.6 ,&\  +0.3  \\
LUB &\ \  a\ \ \ &\  26&\ \ \ \ o\ \ &\ \ \ \ 2&  0.13  &  0.&095 & 1.0   &--3.0 ,&\ --0.6  \\
MAS &\ \  c\ \ \ &\  39&\ \ \ \ s\ \ &\ \ \ \ 1&  0.12  &--0.&040 & 0.630 &--1.0 ,&\  +0.2  \\
MCW &\ \  a\ \ \ &\  62&\ \ \ \ o\ \ &\ \ \ \ 1&  0.09  &--0.&062 & 1.0   &--0.7 ,&\  +0.2  \\
MGN &\ \  a\ \ \ &\  18&\ \ \ \ o\ \ &\ \ \ \ 2&  0.08  &--0.&256 & 1.0   &--3.0 ,&\ --1.1  \\
NHS &\ \  c\ \ \ &\  33&\ \ \ \ o\ \ &\ \ \ \ 2&  0.15  &  0.&070 & 1.0   &--2.6 ,&\ --1.0  \\
OIN &\ \  d\ \ \ &\  15&\ \ \ \ n\ \ &\ \ \ \ 2&  0.114 &  0.&0   & 1.0   &--0.4 ,&\  +0.4  \\
PET &\ \  a\ \ \ &\  28&\ \ \ \ n\ \ &\ \ \ \ 2&  0.15  &  0.&0   & 1.0   &--2.7 ,&\ --0.5  \\
PSB &\ \  a\ \ \ &\  26&\ \ \ \ n\ \ &\ \ \ \ 1&  0.11  &  0.&0   & 1.0   &--3.2 ,&\ --0.7  \\
PSK &\ \  a\ \ \ &\  35&\ \ \ \ o\ \ &\ \ \ \ 2&  0.15  &  0.&0   & 1.0   &--3.0 ,&\ --0.9  \\
RBM &\ \  a\ \ \ &\  32&\ \ \ \ o\ \ &\ \ \ \ 2&  0.17  &--0.&080 & 1.0   &--2.7 ,&\  +0.4  \\
SAH &\ \  c\ \ \ &\  19&\ \ \ \ o\ \ &\ \ \ \ 2&  0.082 &--0.&080 & 1.0   &--0.9 ,&\  +0.3  \\
SIC &\ \ ab\ \ \ &\  20&\ \ \ \ n\ \ &\ \ \ \ 2&  0.15  &  0.&0   & 1.0   &--1.9 ,&\  +0.6  \\
THE &\ \  a\ \ \ &\  12&\ \ \ \ n\ \ &\ \ \ \ 1&  0.13  &  0.&0   & 1.0   &--2.9 ,&\  +0.4  \\
TID &\ \  a\ \ \ &\  35&\ \ \ \ o\ \ &\ \ \ \ 1&  0.08  &  0.&16  & 1.0   &--2.6 ,&\  +0.0  \\
TLA &\ \  a\ \ \ &\  27&\ \ \ \ o\ \ &\ \ \ \ 2&  0.13  &--0.&090 & 1.0   &--2.0 ,&\  +0.0  \\
TLL &\ \  a\ \ \ &\  31&\ \ \ \ o\ \ &\ \ \ \ 2&  0.10  &--0.&129 & 1.0   &--3.0 ,&\ --1.0  \\
WAL &\ \  b\ \ \ &\  27&\ \ \ \ o\ \ &\ \ \ \ 2&  0.16  &  0.&100 & 1.0   &--2.6 ,&\  +0.4  \\
WOR &\ \  a\ \ \ &\ 182&\ \ \ \ n\ \ &\ \ \ \ 2&  0.18  &  0.&0   & 1.0   &--2.7 ,&\  +0.5  \\
ZAS &\ \  c\ \ \ &\  63&\ \ \ \ s\ \ &\ \ \ \ 2&  0.15  &--0.&071 & 0.604 &--1.1 ,&\  +0.2  \\
\hline                                          
\end{tabular}                                                                                      
}
\end{table}
\end{scriptsize}

Once the compilation was finished, we carried out the iterative
procedure performed in CEN01b to end up with a homogeneous system of
stellar atmospheric parameters. 

Most original sources giving any of the three atmospheric parameters
for MILES stars were calibrated and bootstrapped against the reference
system making use of all stars in common between both samples. This
was done separately for each of the three atmospheric parameters (when
available) by comparing the parameter values provided by a certain
source ($p$) against those in the reference system ($p_{\rm ref}$) for
the common subsample of stars. The resulting trends were quantified by
fitting both a linear relationship ($p = A + B\,p_{\rm ref}$) and a
constant offset ($p = A + p_{\rm ref}$). Using a $t$-test and a
significance level of $\alpha = 0.1$, we checked the significance of
the derived fits, that is, whether $B$ --for the linear fit-- and $A$
--for the offset fit-- were significantly different from 1 and 0
respectively. If only one of the two fits was significant, we adopted
it to bootstrap the data from the source against the reference
system. In case that both fits turned out to be significant, we
prefered to keep and apply the linear correction. Obviously, when none
of the fits were statistically significant, the one-to-one
relationship could be assumed and the original parameters were kept.

To ensure that comparisons between any source and the reference system
were statistically significant, a first iteration of the above
procedure was carried out for all those sources that had, at least, 25
stars in common with the complete sample of SKC for any of the three
atmospheric parameters. All the stars whose parameters were coming
from references calibrated and corrected in this way constitute a new
category of reference stars we refer to as RF1. They were added to the
original reference system, thus enlarging the initial sample SKC and
constituting a new, larger reference system (SKC \& RF1).  The whole
above procedure was thus repeated for the rest of original sources
using SKC \& RF1 as reference system. Since, in general, the number of
stars in the remaining, non-calibrated sources was small, the minimum
number of stars in common with the reference system required to
calibrate a given source was set to be only 15. This lead to a second
set of final parameters which is called RF2. We did not perform
further iterations since those sources that had not been calibrated
yet did not possess enough stars in common with the new reference
system (SKC \& RF1 \& RF2) to ensure reliable calibrations.  

Finally, for each of the three atmospheric parameters, an estimation
of the quality of the distinct calibrated sources was determined by
computing the r.m.s. standard deviation of the {\sl corrected}
parameter values w.r.t. those given in the reference system for all
the stars in common. As it will be explained later, weighted according
to the data quality, the various, corrected data sources for each
single star were averaged to provide a final homogeneous set of
atmospheric measurements.

It is important to note that, since the original reference system of
the whole iterative procedure is SKC, all the above calibrations will
in principle be valid for stars within the $T_{\rm eff}$ range spanned
by that work, i.e. from 4000\,K to 6300\,K. In turn, we did not follow
a fully automatic approach and the original parameters for every star
were checked for inconsistencies or outliers, removing original
references when necessary. Note also that, since the procedure is
separately carried out for $T_{\rm eff}$, $\log g$ and [Fe/H], the
number of stars within the subsequent reference categories (SKC \& RF1
and SKC \& RF1 \& RF2) for each of the three parameters does not have
to be necessarily the same.

In Tables~\ref{caltemp}, \ref{calgrav} and \ref{calmetal} we present,
respectively, the details of the calibrations on $T_{\rm eff}$, $\log
g$ and [Fe/H] for all the calibrated sources, with reference codes for
these sources being given in Table~\ref{calrefer}.  The above tables
also include a code indicating the different methods used in each
original paper to derive the atmospheric parameters. Note that,
although the tabulated, r.m.s. standard deviations ($\sigma$) are due
to uncertainties both in the SKC parameters and in the calibrated
reference, a relative comparison of the different values could in
principle provide an estimate of the reliability of the different
methods. Even though a critical analysis of these techniques is out of
the scope of this paper, it must be noted that we do not find any
systematic trend when comparing the uncertainties ($\sigma$) or the
calibration parameters (the derived slopes and independent terms) of
the different working methods.

\begin{scriptsize}
\begin{table*}
\centering{
\caption{Codes for calibrated original references in
Tables~\ref{caltemp}, \ref{calgrav} and \ref{calmetal}.}
\label{calrefer}                                                            
\begin{tabular}{@{}llllll@{}}          
\hline
 Code& Reference &  Code& Reference &  Code& Reference  \\
\hline                                
AAM & Alonso, Arribas \& Mart\'{\i}nez-Roger (1996a)   & GGR & Gratton et al.~(1982)             & MGN & Magain (1989)\\
AFG & Axer, Fuhrmann \& Geheren (1994)                 & GRJ & Gray \& Johanson (1991)           & NHS & Nissen, Hoeg \& Schuster (1997)\\
BAL & Balachandran (1990)                              & GRO & Gratton \& Ortolani (1984)        & OIN & Oinas (1974)\\
BKP & Beers et al.~(1990)                              & GRS & Gratton \& Sneden (1987)          & ONS & O'Neal, Neff \& Saar (1998)\\
BLG & Blackwell \& Lynas-Gray (1994)                   & GRT & Gratton (1989)                    & PET & Peterson (1981)\\
BOV & Bohm-Vitense (1992)                              & GSC & Gratton et al.~(2000)             & PSB & Pilachowski, Sneden \& Booth (1993)\\
BSL & Brown et al.~(1989)                              & HEA & Hearnshaw (1974)                  & PSK & Pilachowski, Sneden \& Kraft (1996)\\
CGC & Carretta et al.~(2000)                           & HWA & Helfer \& Wallerstein (1968)      & RBM & Rebolo, Beckman \& Molaro (1988)\\
CLG & Clementini et al.~(1999)                         & JON & Jones (1997)                      & SAH & Saxner \& Hammarback (1985)\\
CLL & Carney et al.~(1994)                             & KNK & Kunzli et al.~(1997)              & SIC & Silva \& Cornell (1992)\\
CNZ & Chen et al.~(2000)                               & KSP & Krishnaswamy-Gilroy et al.~(1988) & TAY & Taylor (1994)\\
CTL & Clegg, Tomkin \& Lambert (1981)                  & LAI & Laird (1985)                      & THE & Th\'evenin (1998)\\
CTS & Cottrell \& Sneden (1986)                        & LBO & Luck \& Bond (1985)               & TID & Th\'evenin \& Idiart (1999)\\
EAG & Edvardsson et al.~(1993)                         & LCH & Luck \& Challener (1995)          & TLA & Tomkin \& Lambert (1999)\\
FHR & Fuhrmann (1998)                                  & LRS & Lambert \& Ries (1981)            & TLL & Tomkin et al.~(1992)\\
FLB & Fulbright (2000)                                 & LUB & Luck \& Bond (1983)               & WAL & Wallerstein (1962)\\
FRA & Francois (1986)                                  & MAM & Malagnini \& Morossi (1990)       & WOR & Worthey et al.~(1994)\\
FRC & Fern\'andez-Villaca\~nas, Rego \& Cornide (1990) & MAS & Marsakov \& Shevelev (1995)       & ZAS & Zakhozhaj \& Shaparenko (1996)\\
FUH & Fuhrmann (2000)                                  & MCW & McWilliam (1990)                  & & \\
GCC & Gratton, Carretta \& Castelli (1996)             & MEH & Meyer et al.~(1998)               & & \\
\hline                 
\end{tabular}                                             
}               
\end{table*}                                                                
\end{scriptsize}

\subsection{Final atmospheric parameters for field stars}
\label{calculo}

As in CEN01b, the final set of atmospheric parameters for field stars
has been derived in different ways depending on the original
literature sources which were available in each
case. Table~\ref{paramfield} lists the final derived atmospheric
parameters for all the field stars in MILES. A synthesized recipe of
the different approaches and resulting parameter categories is given
in Table~\ref{comentrefs}. A more detailed explanation follows below:

\begin{itemize}

\item If the star is included in the original, reference sample of
SKC, the three atmospheric parameters from that paper were kept (coded
SKC). This turned out to be the case for 164 stars of our sample.

\item When the star is not included in the sample of SKC but in
$N$ previously calibrated sources, and the original parameters are
within the calibration ranges listed in Tables~\ref{caltemp},
\ref{calgrav} and~\ref{calmetal}, the new parameters $P$ were
determined by taking the weighted average:

\begin{equation}P = \frac{\sum_{i=1}^{N} p^{*}_{i}/ \sigma_{i}^{2}}
{\sum_{i=1}^{N} 1/\sigma_{i}^{2}} 
\end{equation} 
where $p^{\ast}_{i}$ is the corrected parameter and {\large
$\sigma$}$_{i}$ corresponds to the r.m.s. standard deviation of the
comparison with the reference system (SKC or RF1 \& SKC; see
Tables~\ref{caltemp}, \ref{calgrav} and~\ref{calmetal}).  It is
important to note that, when the applied correction was either an
offset or a linear relation with slope $\sim 1$, we allowed small
extrapolations of the derived fits to obtain the parameters of stars
slightly out of the validity regime. Most (see below) of the
atmospheric parameters of the stellar library presented here have been
derived in this way (coded RF1 and RF2).

\item When the star is not included in any calibrated source (or, if
included, the atmospheric parameters are well out of the calibration
range), the final parameter is the raw mean value from all the
available original sources and no previous correction to the parameter
value has been applied. 

Unlike CEN01b, in this paper we prefered not to make any category
distinction on the basis of the $T_{\rm eff}$ of the star
whose parameters were derived in this way, so this category is unique
and coded RF3. Obviously, the final parameters of stars within RF3 are
expected to be less reliable than those obtained from calibrated
sources. Since their absolute uncertainties are known to depend on the
$T_{\rm eff}$ regime (parameter determinations are in general
more reliable for intermediate temperatures stars than for early and
late spectral types), an estimation of relative errors for different
temperature regimes is given below.

\item If there is no available data in the literature, both $T_{\rm
eff}$ and $\log g$ are estimated from the spectral type and the
luminosity class using the tabulated atmospheric data from
Lang~(1991). Only a few parameters (0.7 per cent of the temperature
estimations and 1.8 per cent for gravities) were derived in this way,
which we coded as RF6 to follow the notation in CEN01b.

\end{itemize}

With the aim of checking our results and detecting inconsistencies
between stellar spectra and their assigned, final parameters, we
compared the spectrum of every single star with an average one
resulting from the interpolation of the rest of stars in MILES at
exactly the same stellar parameters as those of the problem star. In
order to do this we employed the interpolator code described in
VAZ03. This allowed us to find a few stars whose spectral types were
not compatible with their assigned parameters, the ones were just
removed from the table if no apparent reason was found to drive the
observed discrepancy.

\begin{figure}
\epsfig{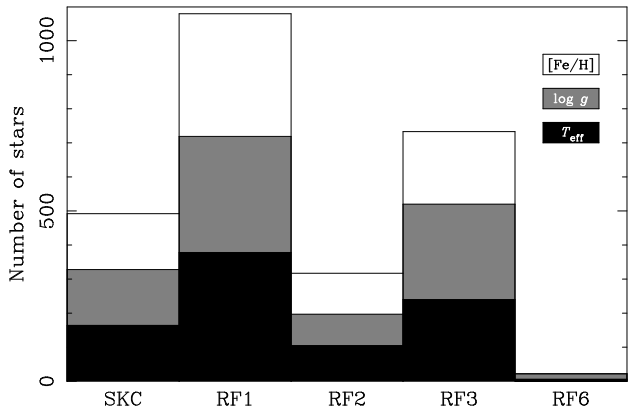}
\caption{Histogram illustrating the total number of stars with
effective temperature (black), gravity (gray) and metallicity (white)
in each category.}
\label{histog}
\end{figure}

To summarize, Figure~\ref{histog} illustrates the number of stars with
final atmospheric parameters in each different category. A total of
893 temperatures, 893 gravities and 857 metallicities were derived for
the 896 field stars of the stellar library. It is worth noting that
most of the $T_{\rm eff}$ (72.5 per cent), $\log g$ (66.9 per cent)
and [Fe/H] (75.1 per cent) values were either taken from the initial
reference system (SKC) or derived from calibrated and corrected
original sources (RF1 and RF2).

As far as the uncertainties of the derived parameters are concerned,
we mostly reproduce the values derived in CEN01b for the distinct
categories. We therefore refer the reader to Section~5 in the above
paper for a detailed explanation of the error estimation. In RF1 and
RF2, typical errors of $\sim 60$\,K, 0.2\,dex and 0.1\,dex are derived
for $T_{\rm eff}$, $\log g$ and [Fe/H] respectively. Clearly the
accuracy of the stellar atmospheric parameters in the RF1 and RF2
categories is much higher than those in the RF3 and RF6 categories,
where calibrations have not been applied. Relative uncertainties for
$T_{\rm eff}$ values of stars in RF3 have been measured to be $\sim 2$
per cent and $\sim 5$ per cent for intermediate spectral types
(4000\,K $<T_{\rm eff}<$ 6300\,K) and extreme spectral types ($T_{\rm
eff} > 6300$\,K; $T_{\rm eff} < 4000$\,K) respectively. Also, the fact
that our uncertainties are pretty consistent with those given in
CEN01b proves that we are indeed measuring the intrinsic uncertainties
among different sources and the results are not affected by
small-numbers statistics.

Finally, we compared the parameters that we obtained with those in
CEN01b. For about 200 out of the 343 stars in common there is no
difference, since our parameters are based on the same literature
references.  For most of the other stars the differences are not
large. In the case of effective temperatues, the difference is more
that 200\,K for only 13 stars. For 9 stars $\log g$ is different by
more than 0.3\,dex, and for 12 stars [Fe/H] differs by more than
0.1\,dex. The good agreement is not surprising, since in this paper we
have used the same method as in CEN01b.

A detailed table containing all the original data that were used to
derive the final atmospheric parameters of the stellar library is
available from:\\{\tt http://www.ucm.es/info/Astrof/miles/miles.html}

\begin{scriptsize}
\begin{table}                                                              
\centering{
\caption{Brief explanation for the different methods to derive the atmospheric parameters.}
\label{comentrefs}							       
\begin{tabular}{@{}ll} 	 
\hline 			       
 SKC & From Soubiran et al.~(1998) \\ 
 RF1 & From calibrated and corrected sources onto SKC \\  
 RF2 & From calibrated and corrected sources onto RF1 \& SKC\\  
 RF3 & From non calibrated sources \\ 
 RF6 & From spectral type and luminosity class (Lang 1991) \\ 
\hline 	   
\end{tabular}  					   
}		 
\end{table}								    
\end{scriptsize}

\section{Atmospheric parameters for cluster stars}

The stellar sample presented in this series of papers constitutes an
extension of both the Lick/IDS stellar library (G93; WOR) and the
near-IR CaT stellar library (CEN01a), with 241 and 403 stars in common
respectively. The above two spectral libraries included a large sample
of open and globular cluster stars, some of which have been retained
in the present version. In this section we revise the atmospheric
parameters of these cluster stars and describe the procedures carried
out to derive such parameters. An updated list with recent
determinations of mean metallicity [Fe/H], color excess $E(B-V)$,
apparent visual distance modulus ($m-M$)$_V$, and age for each cluster
is presented in Table~\ref{paramclust}. Also, the final parameters
adopted for each single star are presented in
Table~\ref{paramstarclust}.

\begin{table*}
\caption{Basic data for clusters: reddening $E(B-V)$, apparent visual
distance modulus $(m-M)_V$, age and metallicity [Fe/H]. Key codes of
sources for the above data: BER92 (Bergbusch \& VandenBerg~1992);
CEN02 (Cenarro et al.~2002); CG97 (Carretta \& Gratton 1997); CHA99
(Chaboyer, Green \& Liebert 1999); GR00 = Gratton (2000); GRU02
(Grundahl et al.~2002); GVA98 (Grundahl, Vandenberg \& Andersen 1998);
HAN04 (Hansen et al.~2004); HAR03 (Catalogue of Galactic Globular
Clusters by Harris 1996; revised version of 2003); KH05 (Catalogue
from Kharchenko et al.~2005 and references therein); LMG94 (Loktin,
Matkin \& Gerasimenko 1994; $(m-M)_V$ is derived from the distance and
the reddening); REY01 (Rey et al.~2001); TWA85 (Twarog \& Tyson 1985);
TWA97 (Twarog, Ashman \& Anthony-Twarog 1997); TWA99 (Twarog,
Anthony-Twarog \& Bricker 1999); VAN83 (VandenBerg~1983); VAN00
(VandenBerg~2000). Key codes for sources of photometric data: A\&H71
(Arp \& Hartwick~1971); A\&L80 (Alcaino \& Liller~1980); B\&S58
(Burbidge \& Sandage~1958); CAR82 (Carney 1982); CLE94 (Clementini et
al.~1994b); CUD85 (Cudworth~1985); FPC83 (Frogel, Persson \&
Cohen~1983); G93 (Gorgas et al.~1993 and references therein. $V$
magnitudes were transformed from $M_V$ and $(m-M)$ as given in
Tables~3 and 4 of that paper. $(V-K)$ data were only considered when
not interpolated from $B-V$; see caption of Table~3 in that paper);
MCN80 (McNamara~1980); MOR78 (Morel \& Magnenat 1978); S\&H77
(Stetson \& Harris~1977); STE03 (Stetson et al.~2003); WOR (Worthey et
al.~1994 and references therein; Table~A2B in that paper); WEB-codes
taken from the "webda" database of Open Clusters at {\tt
http://www.univie.ac.at/webda/} (0014: Johnson \& Knuckles~1955; 0106:
McClure, Forrester \& Gibson~1974; 0191: Gieren~1981; 0312:
Stauffer~1982; 1091: Johnson et al.~1966).}
\label{paramclust}
\begin{center}
\begin{tabular}{llcccccl}
\hline
\multicolumn{2}{l}{Cluster names} &  Type & $E(B-V)$ & ($m - M$)$_V$ & Age (Gyr) & [Fe/H] & Photometric sources \\
\hline
 Alpha Per     & Mel\,20       & open & 0.09 & 6.67 & 0.04 & $-0.05$ & $V$ and $(B-V)$ from WEB-1091 \\
               &               &      & KH05 & KH05 & KH05 & GR00    & no $(V-K)$                    \\

 Coma Ber      & Mel\,111      & open & 0.01   & 4.62  & 0.49  & $-0.05$ & $V$ and $(B-V)$ from WEB-0014\\
               &               &      & LMG94  & LMG94 & LMG94 & GR00    & no $(V-K)$                   \\

 Hyades        & Mel\,25       & open & 0.01   & 3.44   & 0.63   & $+0.13$ &  $V$ and $(B-V)$ from WEB-0014\\
               &               &      & LMG94  & LMG94  &  LMG94 & GR00    &  $(V-K)$ from CAR82 and MOR78 \\

 Pleiades      & Mel\,22       & open & 0.02 & 5.63 & 0.12 & $-0.03$ & V and (B-V) from WEB-0312\\
               &               &      & KH05 & KH05 & KH05 & GR00    & no $(V-K)$               \\

 M3            & NGC\,5272     & globular & 0.01  & 15.12 & 10.30 & $-1.34$ & $V$ and $(B-V)$ from G93 \\
               &               &          & HAR03 & HAR03 & REY01 & CG97    & $(V-K)$ from G93         \\

 M4            & NGC\,6121     & globular & 0.36  & 12.51 & 12.10 & $-1.19$ & $V$ and $(B-V)$ from CLE94 \\
               &               &          & HAR03 & HAN04 & HAN04 & CG97    & no $(V-K)$                 \\

 M5            & NGC\,5904     & globular & 0.03  & 14.46 & 15.00 & $-1.11$ & $V$ and $(B-V)$ from G93 and WOR \\
               &               &          & HAR03 & HAR03 & VAN83 & CG97    & $(V-K)$ from G93                 \\

 M13           & NGC\,6205     & globular & 0.02  & 14.44 & 12.00 & $-1.39$ & $V$ and $(B-V)$ from G93 \\
               &               &          & GVA98 & GVA98 & GVA98 & CG97    & $(V-K)$ from G93         \\

 M25           & IC\,4725      & open & 0.45 & 10.36 & 0.07 & $+0.17$ & $V$ and $(B-V)$ from WEB-0191 \\
               &               &      & KH05 & KH05  & KH05 & GR00    & no $(V-K)$                    \\

 M67           & NGC\,2682     & open & 0.06 & 9.98 & 2.57 & $+0.02$ & $V$ and $(B-V)$ from G93 \\
               &               &      & KH05 & KH05 & KH05 &  GR00   & $(V-K)$ from G93         \\

 M71           & NGC\,6838     & globular & 0.27  & 13.71 & 12.00 & $-0.84$ & $V$ and $(B-V)$ from G93, CUD85 and A\&H71 \\ 
               &               &          & GRU02 & GRU02 & GRU02 & CEN02   & $(V-K)$ from G93                       \\

 M79           & NGC\,1904     & globular & 0.01  & 15.59 & 16.00 & $-1.37$ & $V$ and $(B-V)$ from S\&H77 \\
               &               &          & HAR03 & HAR03 & BER92 & CG97    & $(V-K)$ from FPC83          \\

 M92           & NGC\,6341     & globular & 0.02  & 14.64 & 14.00 & $-2.16$ & $V$ and $(B-V)$ from G93 \\
               &               &          & HAR03 & HAR03 & VAN00 & CG97    & $(V-K)$ from G93         \\

 NGC\,288      & Mel\,3        & globular & 0.03  & 14.95 & 11.51 & $-1.07$ &  $V$ and $(B-V)$ from A\&L80 \\
               &               &          & HAR03 & VAN00 & VAN00 & CG97    &  $(V-K)$ from FPC83          \\

 NGC\,2420     & Mel\,69       & open & 0.05  & 12.31 & 1.90  & $-0.44$ & $V$ and $(B-V)$ from WEB-0106 \\
               &               &      & TWA97 & TWA99 & TWA99 & GR00    & no $(V-K)$                    \\

 NGC\,6791     &               & open & 0.10  & 13.68 & 8.00  & $+0.40$ & $V$ and $(B-V)$ from STE03 \\
               &               &      & CHA99 & CHA99 & CHA99 & GR00    & no $(V-K)$                       \\

 NGC\,7789     & Mel\,245      & open & 0.31  & 12.30 & 1.50  & $-0.13$ & $V$ and $(B-V)$ from MCN80 and B\&S58 \\
               &               &      & TWA85 & TWA85 & TWA85 & GR00    & no $(V-K)$                       \\
\hline
\end{tabular}
\end{center}
\end{table*}

\subsection{Metallicity scale}
\label{metalcum}

Following the criteria adopted for the CAT library, rather than using
the Zinn \& West~(1984; hereafter ZW84) metallicity scale employed in
G93, the metallicity scale of the globular clusters has been
established to be the one defined by Carretta \& Gratton 1997
(hereafter CG97; see also Rutledge, Hesser \& Stetson 1997). The
difference between both scales is specially important at the
intermediate metallicity regime, where the ZW84 scale underestimates
the metallicities as compared to the CG97 system for up to $\sim
0.3$\,dex (e.g. M3, M5). The contrary occurs for the most metal-rich
clusters, as in this case ZW84 metallicities are $\sim 0.1$\,dex
larger than CG97 ones. In this sense, a very interesting case is that
of M71, the globular cluster with the highest [Fe/H] of our sample. On
the basis of the Ca\,{\sc ii} triplet strength, CEN01b discussed that,
in general, the departure of the CaT indices of globular cluster stars
from the index values predicted by the fitting functions derived in
Cenarro et al.~(2002; herafter CEN02) were significantly reduced when
using the CG97 scale instead of the ZW84 one. Even so, CEN02 still
reported the existence of negative CaT residuals for M71 stars that
could only be reasonably explained if their metallicities were lower
($-0.84 \pm 0.06$\,dex) than given by CG97 ($-0.70$\,dex). The last
result is supported by recent spectroscopic determinations of the
metallicity of this cluster: $-0.79 \pm 0.04$\,dex (Sneden et
al.~1994), between $-0.90$ and $-0.75$\,dex (Grundahl et al.~2002),
$-0.80 \pm 0.05$\,dex (King, Boesgaard \& Deliyannis 2005), $-0.80 \pm
0.06$ (Boesgaard et al.~2005), so we decided to keep [Fe/H]$ =
-0.84$\,dex from CEN02 as a reliable value of the metallicity of M71.

Concerning the open clusters in our sample, we have adopted the
metallicity scale constructed by Gratton~(2000; GR00) as it also
relies on the basis of high resolution spectroscopy. In that work,
making use of a compilation of open clusters with metallicities
determined from different techniques, abundances derived from
photometric indices and low resolution spectroscopy are recalibrated
and corrected against high dispersion, spectroscopic
determinations. The final abundances are weighted averages of all the
single --corrected-- abundances. We refer the reader to the above
paper for further details on the procedure.

\subsection{Effective temperatures}
\label{teffs}

Because direct determinations of $T_{\rm eff}$ for cluster stars are
not usual, they have been determined following the same procedure
carried out in CEN01b, that is, by using the empirical,
colour-temperature relations for giant, dwarf and subdwarf stars from
Alonso, Arribas \& Mart\'{\i}nez-Roger~(1996b; 1999; hereafter we will
refer to both references as ALO). In particular, relations involving
$(B-V)$ and $(V-K)$ are the ones considered in this paper.

As demonstrated in CEN01b, temperatures derived from the above
$(V-K)$--temperature relations are consistent with the ones established
by reference system of this paper (SKC), whilst those resulting from
the $(B-V)$ relations exhibit a minor offset of 26\,K that has been
applied to correct and bootstrap the predicted data against the
reference system. In this way, the homogeneity and consistency of
effective temperatures within the whole stellar library are
guaranteed. 

Final $T_{\rm eff}$ values for cluster stars were calculated as an
average of the values derived from $(B-V)$ and $(V-K)$ relations. If
$(B-V)$ was only available, the corrected temperature derived from
this colour was kept. Since the above colour-temperature relations in
turn depend on surface gravity and metallicity, a previous estimate of
both parameters was necessary for each star. In this sense, input
metallicities for each star were the ones established in
Table~\ref{paramclust}. Input surface gravities were taken from G93
and WOR for all stars in common with the Lick/IDS library. For the
rest of stars, we primarily made use of the compilation of Cayrel de
Strobel et al.~(1997; e.g. Clementini et al.~1994a for M4 stars). If
no data for surface gravity was available in the literature, a
tentative value was derived from $(B-V)$ and $M_V$ making use of the
tabulated atmospheric data from Lang~(1991).  Sources for input
photometric data are specified in Table~\ref{paramclust}. In all
cases, appropriate reddening corrections were applied using the color
excesses given in that table and assuming an $E(V-K)/E(B-V)$ value of
2.744 (Harris, Wooff \& Rieke 1978).

It is important to note that, since the colour-temperature
calibrations employed here are just defined for a given range of
effective temperatures ($T_{\rm eff} \leq 8000$\,K), temperatures for
a few early spectral types in our star sample had to be obtained by
means of extrapolations of the above relations. For HD109307, in Coma
Berenices, we derive $T_{\rm eff} = 8471$\,K which is in perfect
agreement with the value determined by Boesgaard~(1987), so the former
value was kept. For the Hyades star HD27962 we obtain 9092\,K, which
is well consistent with different determinations compiled by Cayrel de
Strobel et al.~(1997). Finally, as suggested in B\&S58 and latter
confirmed by McNamara~(1980), NGC\,7789~342 is a blue straggler rather
than a horizontal branch star. On the basis of its $(B-V)$ colour
(McNamara~1980) and corresponding luminosity class, both the spectral
type (B9) and temperature ($T_{\rm eff} \sim 10500$\,K) inferred from
Lang~(1991) are consistent with the temperature derived from the
colour-metallicity calibration ($\sim 10000$\,K). We are therefore
confident that the temperatures derived as extrapolations of the
colour-temperature relations are still safe for most cases considered
in this paper. In this sense, the temperature derived in this way for
the horizontal branch star M5 II-53 (9441\,K) was kept instead of the
value given in WOR (10460\,K).

\subsection{Surface gravities}
\label{logg}

\begin{figure}
\epsfig{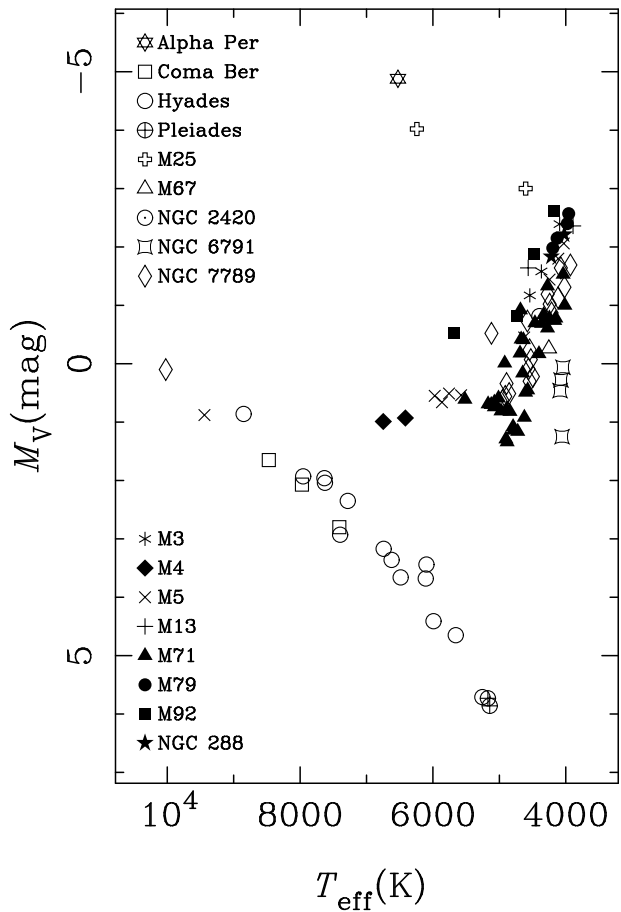}
\caption{Pseudo-HR diagram of cluster stars for which atmospheric
parameters have been computed. Different symbols correspond to stars
in different open (open symbols) and globular (filled and line
symbols) clusters as shown in the top and bottom keys
respectively. Absolute magnitudes are derived from the photometric
data, reddenings and distance moduli given in
Table~\ref{paramclust}. Effective temperatures are computed from
colour-temperature calibrations as explained in Section~\ref{teffs}.}
\label{clustHRdiag}
\end{figure}

Surface gravities for cluster stars were estimated by matching the
location of each star in a $M_V$--$T_{\rm eff}$ diagram to
evolutionary tracks, which is basically the same procedure carried out
by G93 for the Lick/IDS cluster stars. In this paper, however, we use
the improved set of isochrones from Girardi et al.~(2000) after being
transformed to the observational plane (colors and magnitudes) on the
basis of the empirical relations given in ALO (see details in
VAZ03). It is worth noting that the above isochrone set is the same as
employed by our group for stellar population synthesis modeling
(e.g. VAZ03), so the surface gravities derived in this way will be
fully consistent with the values demanded by the spectral synthesis
procedure (Paper~III). Effective temperatures are the ones derived in
Section~\ref{teffs}. Reddening-corrected values of $M_V$ were computed
assuming a Galactic extincion law with $R_V = 3.1$ and deriving the
$V$-band extinction $A_V$. Sources of photometric data, as well as the
adopted age, metallicity [Fe/H], reddening $E(B-V)$, and apparent
distance modulus in the $V$-band $(m-M)_V$ for each cluster are
summarized in Table~\ref{paramclust}.

In Figure~\ref{clustHRdiag}, a pseudo-HR diagram ($M_V$--$T_{\rm
eff}$) for the whole set of cluster stars is presented. Several
additional cluster stars --not finally included in MILES-- have been
considered in the sample. This allows us to determine their
atmospheric parameters in a consistent manner together with the MILES
stars, what may be useful for future work on this topic.

The procedure to determine surface gravities is described below. For
each cluster, taking into account both the age and metallicity values
listed in Table~\ref{paramclust}, we selected from Girardi et
al.~(2000) those two isochrones having the most similar age to that of
the cluster and metallicities enclosing the corresponding value of the
cluster. For each one of the two isochrones, and in order to avoid
uncertainties arising from the colour-temperature relations, a surface
gravity value for each star was estimated by comparing to the
predicted $M_V$, that is, by ignoring any mismatch in $T_{\rm
eff}$. This is a reasonable assumption since relative errors in
absolute magnitudes are expected to be smaller than $T_{\rm eff}$
uncertainties. Final $\log g$ values were computed as weighted means
of the single values derived from the two different metallicity
isochrones, with weights accounting for the distance between the
adopted cluster metallicity and the isochrone values.

\begin{figure*}
\epsfig{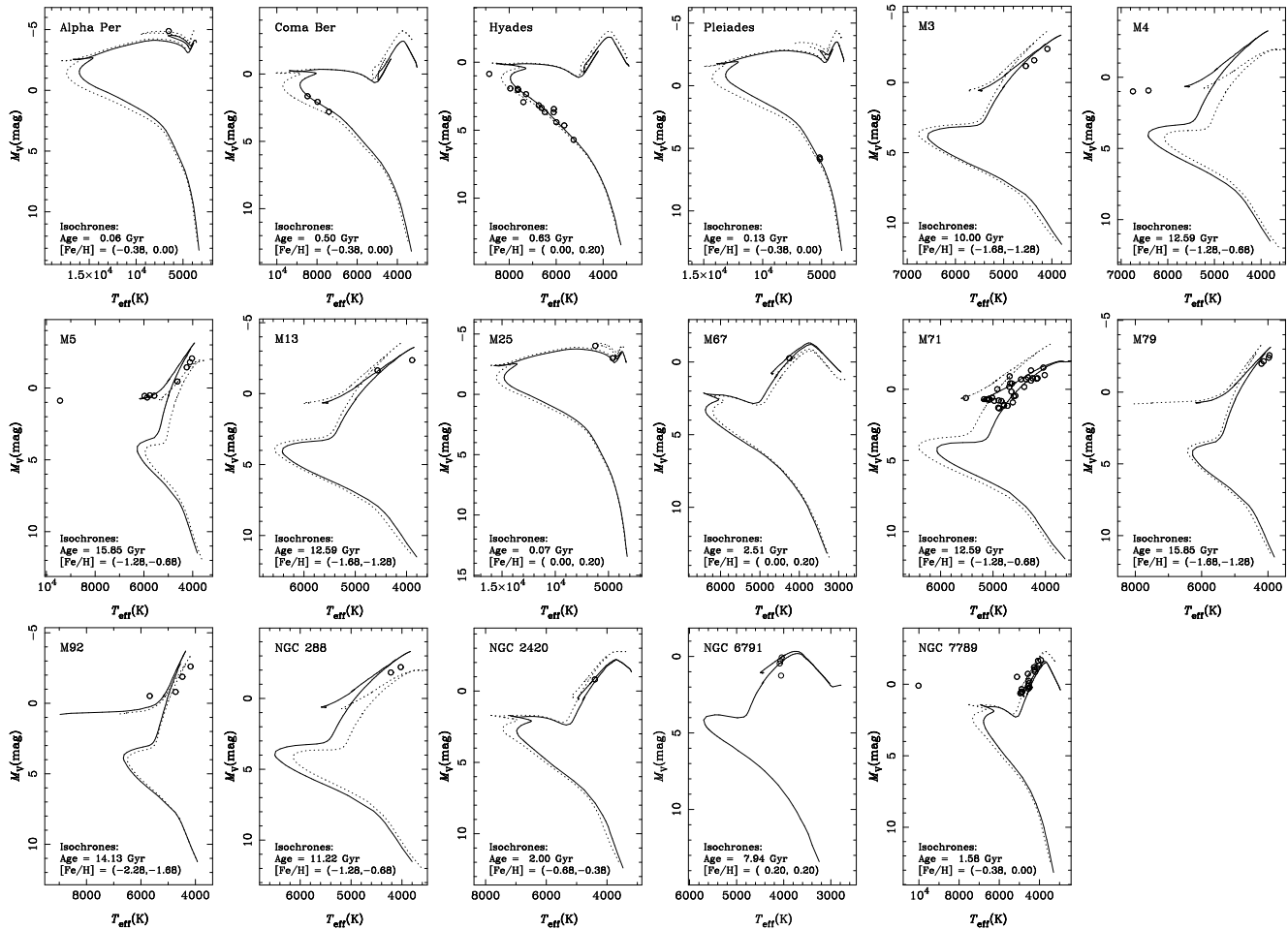}
\caption{Pseudo-HR diagrams for the cluster stars are presented
together with adequate isochrones (Girardi et al.~2000) for each
individual cluster. Open circles are used for individual stars in the
clusters. Solid and dashed lines illustrate isochrones having a
similar age to that of the cluster and two metallicity values
enclosing the one of the cluster (as shown in the labels). Adopted
ages and metallicities for the clusters are given in
Table~\ref{paramclust}. In all cases, the solid line is employed to
indicate the isochrone whose metallicity is closer to that of the
cluster. Since the metallicity adopted for NGC\,6791 ($+0.40$\,dex) is
out of the metallicity regime of Girardi's isochrones, the most
metal-rich one ($+0.20$\,dex) is only displayed. Surface gravity for
each star was estimated by comparing to the predicted $M_V$ as
explained in Section~\ref{logg}.}
\label{isofigs}
\end{figure*}

Figure~\ref{isofigs} illustrates qualitatively the above
procedure. The two isochrones employed for each cluster (with solid
and dashed lines indicating, respectively, those having the higher and
lower weight in the final value of $\log g$) are overplotted together
with the location of the cluster stars in a pseudo-HR diagram
($M_V-T_{\rm eff}$). Overall, the agreement between the isochrones and
stars is reasonably good.

\section{Summary}

\begin{figure}
\epsfig{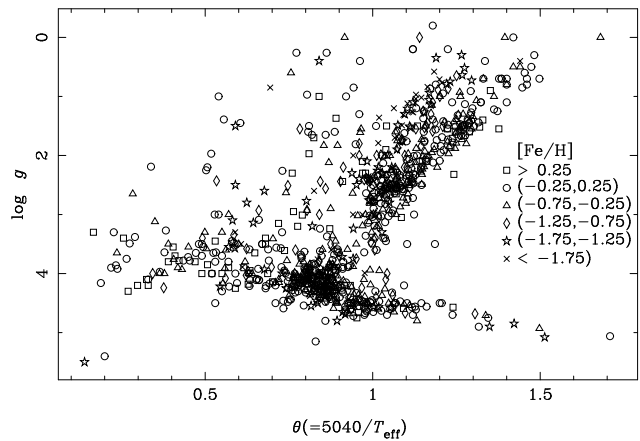}
\caption{Surface gravity versus effective temperature diagram for all
the library stars. Different symbols are used for stars within distinct
metallicity regimes as indicated in the key.}
\label{MILESHR}
\end{figure}

The uncertainties in the input atmospheric parameters of library stars
are one of the main sources of potential errors when computing the
predictions of evolutionary synthesis models. In this paper we have
derived a reliable, and highly homogeneous, set of atmospheric
parameters ($2748 < T_{\rm eff} < 36000$\,K; $0.00 < \log g <
5.50$\,dex; $-2.93 <$ [Fe/H] $< +1.65$\,dex) for the 985 stars in
MILES, a new stellar library in the optical spectral range
(Paper~I). For the subsample of 896 field stars, systematic deviations
between parameters from different sources have been calibrated and
corrected by bootstrapping them onto a reference system, following the
procedure stated in CEN01b. Also, the atmospheric parameters of 89
Galactic cluster stars (from 9 open clusters and 8 globular clusters)
have been computed in a homogeneous way: effective temperatures have
been derived from colour-temperature relations in Alonso, Arribas \&
Mart\'{\i}nez-Roger~(1996b; 1999), and surface gravities have been
computed by fitting the location of each star in a pseudo-HR diagram
using appropriate isochrones from Girardi et
al.~(2000). Figure~\ref{MILESHR} shows the complete stellar library in
the parameter space of $T_{\rm eff}$ and $\log g$ for various
metallicity-ranges. In Paper~III of this series we will make use of
the stellar spectra in Paper~I and the atmospheric parameters here
presented to predict the integrated SEDs of stellar populations all
over the spectral range of MILES. Moreover, the usefulness of the new
set of improved parameters goes beyond the objectives of this
series. In particular, it should represent a basic ingredient for the
new generation of spectral synthesis work as well as to improve the
existing empirical calibrations of other relevant spectral features.

\section*{ACKNOWLEDGMENTS}

We thank the referee Raffaele Gratton for fruitful suggestions that
improved the quality of this paper. This work was supported by the
Spanish research projects AYA2003-01840 and AYA2004-03059. SOS
acknowledges the financial support by the Turkish Scientific and
Technical Research Council (T\"{U}B{\.I}TAK) through the NATO-B2
fellowship and the School of Physics and Astronomy (University of
Nottingham). AV is a Ram\'on y Cajal Fellow of the Spanish Ministry of
Education and Science. This research has made use of the Simbad
database (operated at CDS, Strasbourg, France), the NASA's
Astrophysics Data System Article Service, and the Hipparcos Input
Catalog. This paper is based on observations obtained at the Isaac
Newton Telescope, operated by the Isaac Newton Group in the Spanish
Observatorio del Roque de los Muchachos of the Instituto de
Astrof\'{\i}sica de Canarias. The authors are grateful to the
ASTRON-PC and CAT for generous allocation of telescope time.

\appendix

\section{Atmospheric parameters of MILES stars}

Tables~\ref{paramfield} and \ref{paramstarclust} present the
atmospheric parameters ($T_{\rm eff}$, $\log g$ and [Fe/H]) computed,
respectively, for field and cluster stars in MILES.  Identifying
numbers, spectral types, reddening-corrected absolute magnitudes in
$V$ band (only for cluster stars), reference sources, as well as other
stellar libraries whose stars are in common with MILES, are also
provided.

\begin{scriptsize}
\begin{table*}                                                          
\begin{center}
\caption{Final atmospheric parameters of field stars. MILES ordering
numbers have been assigned on the basis of increasing right ascension
(J2000). The corresponding ordering numbers for all stars in common
with the near-IR, CaT stellar libray (CEN01a) are provided in the
third column. Sources for spectral types are the Bright Star Catalog
(Hoffleit \& Jaschek 1982), Andrillat, Jaschek \& Jaschek (1995),
Gorgas et al.~(1999), the Hipparcos Input Catalog and the Simbad
database at {\tt http://simbad.u-strasbg.fr/Simbad}. References for
atmospheric parameters: SKC from Soubiran et al.~(1998). Numerical
references $ijk$ indicate that $T_{\rm eff}$ is from RF$i$, $\log g$
from RF$j$ and [Fe/H] from RF$k$ (see Table~\ref{comentrefs}). Last
column includes codes for other stellar libraries the stars are in
common with: L (Lick/IDS; G93 and WOR); J (Jones 1997); S (STELIB; Le
Borgne et al.~2003); I (Indo-US; Valdes et al.~2004); E (ELODIE.3;
Prugniel \& Soubiran 2004).}
\label{paramfield}
\begin{tabular}{lcccrrrll}
\hline
HD/Other & MILES Num & CaT Num & SpT & $T_{\rm eff}$ & $\log g$ & [Fe/H] & Refs & Libraries \\
\hline
BD+002058A      & 0272 &     & sd:F        &   6024 &  4.50 & --1.49 & 131 &           \\ 
BD-010306       & 0081 &     & G1V         &   5650 &  4.40 & --0.90 & 111 &           \\ 
BD-011792       & 0266 &     & G6V         &   4948 &  3.05 & --1.05 & 111 &           \\ 
BD+012916       & 0505 & 677 & KIIvw       &   4238 &  0.35 & --1.47 & 221 &       I   \\ 
BD-032525       & 0329 &     & F3          &   5750 &  3.60 & --1.90 & 111 &           \\ 
BD+044551       & 0777 &     & F7Vwe       &   5770 &  3.87 & --1.62 & SKC &       I E \\ 
BD-052678       & 0327 &     & F7          &   5429 &  4.43 & --2.14 & 111 &           \\
\hline
\end{tabular}
\end{center}
\end{table*}
\end{scriptsize}

\begin{scriptsize}
\begin{table*}                                                          
\begin{center}
\contcaption{}
\begin{tabular}{lcccrrrll}
\hline
HD/Other & MILES Num & CaT Num & SpT & $T_{\rm eff}$ & $\log g$ & [Fe/H] & Refs & Libraries \\
\hline
HD000245        & 0008 &     & G2V         &   5348 &  4.50 & --1.16 & SKC &         E \\ 
HD000249        & 0005 & 158 & K1IV        &   4717 &  2.40 & --0.34 & 232 &   J   I   \\ 
HD000319        & 0006 &     & A1V         &   8140 &  3.80 & --0.70 & 333 &           \\ 
HD000400        & 0007 &     & F8IV        &   6205 &  4.12 & --0.33 & SKC &   J   I E \\ 
HD000448        & 0009 &     & G9III       &   4710 &  2.56 &   0.05 & 111 &           \\ 
HD000886        & 0011 &     & B2IV        &  21581 &  3.86 &   0.06 & 333 &       I E \\ 
HD001326B       & 0012 & 691 & M6 V        &   3330 &  5.08 & --1.40 & 111 & L         \\
\hline
\end{tabular}
\end{center}
\end{table*}
\end{scriptsize}

\begin{scriptsize}
\begin{table*}                                                          
\begin{center}
\contcaption{}
\begin{tabular}{lcccrrrll}
\hline
HD/Other & MILES Num & CaT Num & SpT & $T_{\rm eff}$ & $\log g$ & [Fe/H] & Refs & Libraries \\
\hline
HD009562        & 0061 &     & G2IV        &   5728 &  3.77 &   0.15 & 111 &   J   I E \\ 
HD009826        & 0063 & 184 & F8V         &   6134 &  4.09 &   0.11 & 111 &   J   I E \\ 
HD009919        & 0064 &     & F0V         &   6830 &  3.96 & --0.41 & 122 &         E \\ 
HD010307        & 0066 & 185 & G2V         &   5838 &  4.28 &   0.03 & 111 & L J   I E \\ 
HD010380        & 0065 & 186 & K3III       &   4057 &  1.43 & --0.25 & SKC & L       E \\ 
HD010700        & 0067 & 188 & G8V         &   5264 &  4.36 & --0.50 & SKC & L     I E \\ 
HD010780        & 0069 & 189 & K0V         &   5392 &  4.60 &   0.10 & 122 & L J   I E \\
\hline
\end{tabular}
\end{center}
\end{table*}
\end{scriptsize}

\begin{scriptsize}
\begin{table*}                                                          
\begin{center}
\contcaption{}
\begin{tabular}{lcccrrrll}
\hline
HD/Other & MILES Num & CaT Num & SpT & $T_{\rm eff}$ & $\log g$ & [Fe/H] & Refs & Libraries \\
\hline
HD023249        & 0124 & 223 & K0IV        &   4884 &  3.40 & --0.11 & 111 & L J   I   \\ 
HD023261        & 0125 &     & G5          &   5132 &  4.64 &   0.10 & 333 &           \\ 
HD023439A       & 0127 & 224 & K1V         &   4990 &  4.40 & --1.01 & 111 & L         \\ 
HD023439B       & 0128 & 225 & K2V         &   4755 &  4.37 & --1.02 & 111 & L         \\ 
HD023607        & 0129 &     & A7V         &   7875 &  4.00 & --0.30 & 333 &           \\ 
HD023841        & 0130 & 226 & K1III       &   4292 &  1.30 & --0.95 & 231 &   J   I   \\ 
HD023924        & 0131 &     & A7V         &   8000 &  4.00 & --0.30 & 333 &           \\
\hline
\end{tabular}
\end{center}
\end{table*}
\end{scriptsize}

\begin{scriptsize}
\begin{table*}                                                          
\begin{center}
\contcaption{}
\begin{tabular}{lcccrrrll}
\hline
HD/Other & MILES Num & CaT Num & SpT & $T_{\rm eff}$ & $\log g$ & [Fe/H] & Refs & Libraries \\
\hline
HD038392        & 0190 &     & K2V         &   4727 &  4.50 &   0.02 & 133 &           \\ 
HD038393        & 0191 & 251 & F6V         &   6331 &  4.27 & --0.05 & 111 & L         \\ 
HD038545        & 0193 &     & A3Vn        &   8970 &  3.60 & --0.30 & 333 &           \\ 
HD038656        & 0195 & 252 & G8III       &   4928 &  2.52 & --0.22 & 111 &   J   I   \\ 
HD038751        & 0194 & 253 & G8IIIvar    &   4751 &  2.27 &   0.04 & 111 & L         \\ 
HD039364        & 0196 &     & G8III/IV    &   4569 &  2.46 & --0.90 & 111 &           \\ 
HD039801        & 0199 & 255 & M2Iab       &   3550 &  0.00 &   0.03 & 233 &       I   \\
\hline
\end{tabular}
\end{center}
\end{table*}
\end{scriptsize}

\begin{scriptsize}
\begin{table*}                                                          
\begin{center}
\contcaption{}
\begin{tabular}{lcccrrrll}
\hline
HD/Other & MILES Num & CaT Num & SpT & $T_{\rm eff}$ & $\log g$ & [Fe/H] & Refs & Libraries \\
\hline
HD057061        & 0252 & 283 & 09Ib        &  32300 &  3.00 &   ---- & 36  &           \\ 
HD057264        & 0253 & 285 & G8III       &   4620 &  2.72 & --0.33 & 111 &   J   I   \\ 
HD058207        & 0254 & 286 & K0III       &   4788 &  2.55 & --0.12 & 111 & L J   I E \\ 
HD058551        & 0255 & 287 & F6V         &   6165 &  4.18 & --0.55 & 111 &   J S I   \\ 
HD059374        & 0257 &     & F8V         &   5800 &  4.40 & --0.90 & 111 &           \\ 
HD059612        & 0256 & 289 & A5Ib        &   8330 &  1.45 &   0.08 & 333 &           \\ 
HD059881        & 0260 &     & F0III       &   7636 &  3.09 &   0.20 & 331 &   J S I   \\
\hline
\end{tabular}
\end{center}
\end{table*}
\end{scriptsize}

\begin{scriptsize}
\begin{table*}                                                          
\begin{center}
\contcaption{}
\begin{tabular}{lcccrrrll}
\hline
HD/Other & MILES Num & CaT Num & SpT & $T_{\rm eff}$ & $\log g$ & [Fe/H] & Refs & Libraries \\
\hline
HD074442        & 0315 & 318 & K0III       &   4657 &  2.51 & --0.06 & 111 & L         \\ 
HD074462        & 0319 & 319 & G5IV        &   4527 &  1.53 & --1.41 & 111 &       I   \\ 
HD074721        & 0317 &     & A0V         &   8560 &  3.57 & --1.42 & 333 &   J   I   \\ 
HD075318        & 0320 &     & G5          &   5422 &  4.50 & --0.04 & SKC &         E \\ 
HD075691        & 0321 &     & K3III       &   4270 &  2.12 & --0.05 & 111 &           \\ 
HD075732        & 0322 & 320 & G8V         &   5079 &  4.48 &   0.16 & SKC & L J S I E \\ 
HD076151        & 0323 &     & G2V         &   5692 &  4.28 &   0.08 & 111 &   J S I E \\
\hline
\end{tabular}
\end{center}
\end{table*}
\end{scriptsize}

\begin{scriptsize}
\begin{table*}                                                          
\begin{center}
\contcaption{}
\begin{tabular}{lcccrrrll}
\hline
HD/Other & MILES Num & CaT Num & SpT & $T_{\rm eff}$ & $\log g$ & [Fe/H] & Refs & Libraries \\
\hline
HD088986        & 0378 &     & G0V         &   5800 &  4.07 &   0.01 & SKC &   J   I E \\ 
HD089010        & 0379 & 339 & G1.5IV-V    &   5698 &  3.92 &   0.01 & 111 & L J   I E \\ 
HD089254        & 0380 &     & F2III       &   7304 &  3.78 &   0.25 & 331 &   J S I   \\ 
HD089449        & 0381 & 341 & F6IV        &   6493 &  4.06 &   0.21 & 111 & L J   I   \\ 
HD089484        & 0382 &     & K1IIIb      &   4470 &  2.12 & --0.38 & 111 &       I   \\ 
HD089707        & 0383 &     & G1V         &   5957 &  4.37 & --0.41 & 111 &   J   I   \\ 
HD089744        & 0384 &     & F7V         &   6219 &  3.95 &   0.23 & 111 &   J   I E \\
\hline
\end{tabular}
\end{center}
\end{table*}
\end{scriptsize}

\begin{scriptsize}
\begin{table*}                                                          
\begin{center}
\contcaption{}
\begin{tabular}{lcccrrrll}
\hline
HD/Other & MILES Num & CaT Num & SpT & $T_{\rm eff}$ & $\log g$ & [Fe/H] & Refs & Libraries \\
\hline
HD108915        & 0443 &     & K0          &   ---- &  ---- &   ---- &     &           \\ 
HD109443        & 0444 &     & F0V         &   6632 &  4.20 & --0.65 & 111 &           \\ 
HD109871        & 0445 &     & K2          &   ---- &  ---- &   ---- &     &           \\ 
HD109995        & 0446 & 369 & A0 V (HB)   &   8300 &  3.50 & --1.99 & 112 & L J S     \\ 
HD110014        & 0447 &     & K2III       &   4379 &  1.37 &   0.11 & 222 &   J       \\ 
HD110379        & 0448 &     & F0V+?       &   6875 &  4.46 & --0.09 & 111 &           \\ 
HD110885        & 0450 &     & G0          &   5253 &  2.40 & --1.53 & 222 &           \\
\hline
\end{tabular}
\end{center}
\end{table*}
\end{scriptsize}

\begin{scriptsize}
\begin{table*}                                                          
\begin{center}
\contcaption{}
\begin{tabular}{lcccrrrll}
\hline
HD/Other & MILES Num & CaT Num & SpT & $T_{\rm eff}$ & $\log g$ & [Fe/H] & Refs & Libraries \\
\hline
HD124850        & 0502 & 408 & F7 IV       &   6116 &  3.83 & --0.11 & 111 &   J S I E \\ 
HD124897        & 0501 & 409 & K2 IIIp     &   4361 &  1.93 & --0.53 & SKC & L J   I E \\ 
HD125184        & 0503 &     & F9V         &   5536 &  3.92 &   0.15 & 111 &   J     E \\ 
HD125451        & 0504 &     & F5IV        &   6669 &  4.44 &   0.05 & 111 &   J   I   \\ 
HD126053        & 0507 &     & G1V         &   5662 &  4.50 & --0.45 & SKC &         E \\ 
HD126141        & 0506 &     & F5V         &   6748 &  4.48 &   0.04 & 111 &   J S I   \\ 
HD126218        & 0509 &     & K0III       &   4921 &  2.81 &   0.09 & 111 &           \\
\hline
\end{tabular}
\end{center}
\end{table*}
\end{scriptsize}

\begin{scriptsize}
\begin{table*}                                                          
\begin{center}
\contcaption{}
\begin{tabular}{lcccrrrll}
\hline
HD/Other & MILES Num & CaT Num & SpT & $T_{\rm eff}$ & $\log g$ & [Fe/H] & Refs & Libraries \\
\hline
HD138776        & 0562 &     & K0          &   5700 &  4.20 &   0.48 & 333 &           \\ 
HD139195        & 0564 &     & K0p         &   4856 &  2.82 & --0.04 & SKC &       I E \\ 
HD139446        & 0566 &     & G8III/IV    &   5080 &  2.70 & --0.26 & 111 &   J   I   \\ 
HD139641        & 0565 &     & G7.5IIIb    &   5030 &  2.99 & --0.49 & 111 &   J S I   \\ 
HD139669        & 0561 & 446 & K5 III      &   3895 &  1.41 & --0.01 & 111 & L J S     \\ 
HD140160        & 0567 &     & A0p...      &   9164 &  3.30 & --0.25 & 333 &           \\ 
HD140283        & 0568 & 447 & F3 VI       &   5687 &  3.55 & --2.53 & SKC & L J     E \\
\hline
\end{tabular}
\end{center}
\end{table*}
\end{scriptsize}

\begin{scriptsize}
\begin{table*}                                                          
\begin{center}
\contcaption{}
\begin{tabular}{lcccrrrll}
\hline
HD/Other & MILES Num & CaT Num & SpT & $T_{\rm eff}$ & $\log g$ & [Fe/H] & Refs & Libraries \\
\hline
HD156026        & 0625 &     & K5V         &   4541 &  4.54 & --0.37 & 111 &           \\ 
HD156283        & 0624 &     & K3IIvar     &   4163 &  1.42 & --0.07 & 111 &           \\ 
HD157089        & 0628 & 486 & F9 V        &   5785 &  4.12 & --0.56 & SKC & L J     E \\ 
HD157214        & 0627 & 487 & G0 V        &   5682 &  4.25 & --0.39 & SKC & L J S   E \\ 
HD157373        & 0626 &     & F4V         &   6422 &  4.07 & --0.60 & 111 &           \\ 
HD157856        & 0631 &     & F3V         &   6330 &  3.91 & --0.14 & 222 &           \\ 
HD157881        & 0630 & 488 & K7 V        &   4065 &  4.57 &   0.38 & 131 &   J     E \\
\hline
\end{tabular}
\end{center}
\end{table*}
\end{scriptsize}

\begin{scriptsize}
\begin{table*}                                                          
\begin{center}
\contcaption{}
\begin{tabular}{lcccrrrll}
\hline
HD/Other & MILES Num & CaT Num & SpT & $T_{\rm eff}$ & $\log g$ & [Fe/H] & Refs & Libraries \\
\hline
HD171496        & 0683 &     & G8IV        &   4700 &  1.60 & --0.91 & 222 &           \\ 
HD171999        & 0684 &     & G5          &   5031 &  4.65 & --0.10 & 131 &           \\ 
HD172103        & 0686 &     & F1IV-V      &   6418 &  3.68 & --0.22 & 111 &           \\ 
HD172365        & 0687 & 529 & F8 Ib-II    &   5800 &  2.12 & --0.36 & 333 &       I E \\ 
HD172380        & 0685 & 530 & M4-5 II     &   3349 &  0.55 &   ---- & 26  &           \\ 
HD172816        & 0691 & 531 & M5.2 III    &   3270 &  0.50 &   ---- & 36  &       I   \\ 
HD172958        & 0688 & 532 & B8 V        &  11300 &  3.75 &   ---- & 11  & L     I   \\
\hline
\end{tabular}
\end{center}
\end{table*}
\end{scriptsize}

\begin{scriptsize}
\begin{table*}                                                          
\begin{center}
\contcaption{}
\begin{tabular}{lcccrrrll}
\hline
HD/Other & MILES Num & CaT Num & SpT & $T_{\rm eff}$ & $\log g$ & [Fe/H] & Refs & Libraries \\
\hline
HD188512        & 0743 & 577 & G8 IVvar    &   5041 &  3.04 & --0.04 & SKC & L       E \\ 
HD188650        & 0741 &     & Fp          &   5764 &  2.90 & --0.40 & 333 &           \\ 
HD188727        & 0744 & 578 & G5 Ib var   &   5685 &  1.60 &   0.00 & 112 & L         \\ 
HD188947        & 0745 &     & K0IIIvar    &   4829 &  2.53 &   0.01 & SKC &   J   I E \\ 
HD189005        & 0746 &     & G8II/III    &   5060 &  2.78 & --0.38 & 111 &           \\ 
HD189558        & 0747 &     & G0/G1V      &   5786 &  3.85 & --1.19 & SKC &       I E \\ 
HD189849        & 0748 &     & A4III       &   7972 &  3.59 &   0.08 & 331 &           \\
\hline
\end{tabular}
\end{center}
\end{table*}
\end{scriptsize}

\begin{scriptsize}
\begin{table*}                                                          
\begin{center}
\contcaption{}
\begin{tabular}{lcccrrrll}
\hline
HD/Other & MILES Num & CaT Num & SpT & $T_{\rm eff}$ & $\log g$ & [Fe/H] & Refs & Libraries \\
\hline
HD204543        & 0803 &     & G0          &   4617 &  1.31 & --1.76 & SKC &       I E \\ 
HD204587        & 0804 & 613 & M0 V        &   4035 &  4.67 &   ---- & 11  & L         \\ 
HD204613        & 0799 &     & G0IIIwsp    &   5727 &  3.84 & --0.51 & SKC &       I E \\ 
HD204754        & 0802 &     & B8III       &  12923 &  3.50 & --0.28 & 333 &           \\ 
HD205021        & 0800 &     & B2IIIvar    &  26740 &  4.16 & --0.23 & 333 &       I   \\ 
HD205153        & 0806 & 615 & G0 IV       &   5947 &  3.70 & --0.02 & 222 & L         \\ 
HD205435        & 0805 & 616 & G8 III      &   5071 &  2.74 & --0.16 & SKC &   J   I E \\
\hline
\end{tabular}
\end{center}
\end{table*}
\end{scriptsize}

\begin{scriptsize}
\begin{table*}                                                          
\begin{center}
\contcaption{}
\begin{tabular}{lcccrrrll}
\hline
HD/Other & MILES Num & CaT Num & SpT & $T_{\rm eff}$ & $\log g$ & [Fe/H] & Refs & Libraries \\
\hline
HD218235        & 0858 &     & F6Vs        &   6490 &  4.08 &   0.16 & 111 &           \\ 
HD218329        & 0859 & 653 & M1 IIIab    &   3810 &  1.10 &   ---- & 11  & L       E \\ 
HD218502        & 0860 & 655 & F3w         &   6030 &  3.76 & --1.84 & SKC &       I E \\ 
HD218640        & 0861 &     & A3IV:       &   5640 &  3.62 &   0.27 & 111 &           \\ 
HD218804        & 0862 &     & F5IV        &   6261 &  4.05 & --0.23 & 222 &         E \\ 
HD218857        & 0863 & 657 & G6w         &   5082 &  2.41 & --1.93 & SKC &       I E \\ 
HD219116        & 0865 &     & G8IIIp      &   5300 &  3.50 & --0.34 & 333 &           \\
\hline
\end{tabular}
\end{center}
\end{table*}
\end{scriptsize}

\begin{scriptsize}
\begin{table*}                                                          
\begin{center}
\caption{Final atmospheric parameters ($T_{\rm eff}$, $\log g$ and
[Fe/H]) and absolute magnitudes in $V$-band ($M_V^0$) of cluster
stars. References for effective temperatures: (1) Derived from $B-V$
versus $T_{\rm eff}$ relations in ALO (Alonso, Arribas \&
Mart\'{\i}nez-Roger~1996b, 1999); (2) Mean value from $B-V$ and $V-K$
versus $T_{\rm eff}$ relations in ALO. Surface gravities were derived
by interpolating the location of each individual star in a $M_V-T_{\rm
eff}$ diagram, using appropriate isochrones taken from Girardi et
al.~(2000); see Section~\ref{logg}. Metallicity sources are provided
in Table~\ref{paramclust}. Spectral types are presented for stars in
Alpha Per, Coma Ber, Hyades, Pleiades, M25 and NGC\,6791. For the rest
of clusters, we list positions in the HR diagram (SGB: subgiant
branch; GB: giant branch; HB: horizontal branch; AGB: asymptotic giant
branch). Stars labeled with an asterisk were not finally included in
the MILES spectral database. Those stars being in common with the
Lick/IDS sample are coded as 'L' in the last column.}
\label{paramstarclust}
\begin{tabular}{llcclrrrrcc}
\hline
Cluster name  & Star ID & MILES Num & CaT Num  & SpT     &$M_V^0$ & $T_{\rm eff}$ & $\log g$ & [Fe/H] & Ref($T_{\rm eff}$) & Library \\
\hline
Alpha Per     & HD 020902         & 0898  &       & F5 Ib   &$-4.88 $& 6527. & 0.26 &$-0.05$& 1 &  \\
      
Coma Ber      & HD 107276         & 0920  &  006  & A6 IV-V &$ 2.07 $& 7972. & 4.21 &$-0.05$& 1 &L \\
              & HD 107513         & 0921  &  007  & A9 V    &$ 2.80 $& 7409. & 4.25 &$-0.05$& 1 &L \\
              & HD 109307         & 0922  &       & A4 V    &$ 1.65 $& 8471. & 4.16 &$-0.05$& 1 &  \\
      
Hyades        & HD 025825         & 0901  &  016  & G0 V    &$ 4.41 $& 5992. & 4.41 &$+0.13$& 1 &L \\
              & HD 026736         & 0902  &  017  & G3 V    &$ 4.65 $& 5657. & 4.45 &$+0.13$& 2 &L \\
              & HD 027383         & 0904  &       & F7 V+?  &$ 3.44 $& 6098. & 4.28 &$+0.13$& 2 &  \\
              & HD 027524         & 0905  &  023  & F5 V    &$ 3.36 $& 6622. & 4.28 &$+0.13$& 1 &L \\
              & HD 027561         & 0906  &  025  & F5 V    &$ 3.17 $& 6742. & 4.24 &$+0.13$& 2 &L \\
              & HD 027962         & 0907  &       & A2IV    &$ 0.86 $& 8850. & 3.80 &$+0.13$& 2 &  \\
              & HD 028483         & 0908  &  029  & F6 V    &$ 3.66 $& 6486. & 4.30 &$+0.13$& 1 &L \\
              & HD 028546         & 0909  &       &   Am    &$ 2.04 $& 7626. & 4.11 &$+0.13$& 1 &  \\
              & HD 029375         & 0910  &  032  & F0 V    &$ 2.35 $& 7283. & 4.17 &$+0.13$& 1 &L \\
              & HD 030034         & 0911  &  034  & F0 V    &$ 1.96 $& 7634. & 4.11 &$+0.13$& 1 &L \\
              & HD 030210         & 0912  &  035  &   Am    &$ 1.93 $& 7954. & 4.11 &$+0.13$& 1 &L \\
              & HD 030676         & 0913  &       &   F8    &$ 3.68 $& 6108. & 4.30 &$+0.13$& 1 &  \\
              & HD 031236         & 0914  &  036  & F3 IV   &$ 2.93 $& 7397. & 4.21 &$+0.13$& 1 &L \\
              & HD 284253         & 0903  &  020  & K0 V    &$ 5.71 $& 5256. & 4.55 &$+0.13$& 1 &L \\
      
Pleiades      & Mel\,22 0296      & 0899  &       &  G8V    &$ 5.73 $& 5173. & 4.57 &$-0.03$& 1 &  \\
              & Mel\,22 2462      & 0900  &       &         &$ 5.86 $& 5145. & 4.61 &$-0.03$& 1 &  \\
      
M3            & M3 398            & 0925  &  043  &  GB     &$-1.17 $& 4541. & 1.51 &$-1.34$& 1 &L \\
              & M3 III 28         & 0924  &  044  &  GB     &$-2.39 $& 4093. & 0.75 &$-1.34$& 2 &L \\
              & M3 IV 25          & 0923  &  045  &  GB     &$-1.58 $& 4367. & 1.27 &$-1.34$& 2 &L \\
      
M4            & M4 LEE 1409       &  (*)  &       &         &$ 0.93 $& 6415. & 2.51 &$-1.19$& 1 &  \\
              & M4 LEE 2303       & 0933  &       &         &$ 0.99 $& 6748. & 2.51 &$-1.19$& 1 &  \\
      
M5            & M5 I 45           &  (*)  &  046  &  HB     &$ 0.51 $& 5758. & 2.40 &$-1.11$& 1 &L \\
              & M5 II 51          & 0927  &  047  &  GB     &$-0.45 $& 4627. & 1.74 &$-1.11$& 2 &L \\
              & M5 II 53          & 0929  &  048  &  HB     &$ 0.88 $& 9441. & 2.43 &$-1.11$& 1 &L \\
              & M5 II 76          & 0928  &  049  &  HB     &$ 0.55 $& 5974. & 2.44 &$-1.11$& 1 &L \\
              & M5 III 03         & 0926  &  050  &  GB     &$-2.06 $& 4031. & 0.65 &$-1.11$& 2 &L \\
              & M5 IV 19          & 0930  &  051  &  GB     &$-1.80 $& 4113. & 0.87 &$-1.11$& 2 &L \\
              & M5 IV 59          &  (*)  &  052  &  GB     &$-1.44 $& 4245. & 1.15 &$-1.11$& 2 &L \\
              & M5 IV 86          & 0931  &  053  &  HB     &$ 0.54 $& 5576. & 2.44 &$-1.11$& 2 &L \\
              & M5 IV 87          & 0932  &  054  &  HB     &$ 0.66 $& 5864. & 2.56 &$-1.11$& 1 &L \\
      
M13           & M13 A 171         & 0934  &  040  &  AGB    &$-1.64 $& 4566. & 1.21 &$-1.39$& 1 &L \\
              & M13 B 786         & 0935  &  041  &  GB     &$-2.36 $& 3891. & 0.73 &$-1.39$& 1 &L \\
                  
M25           & HD 170764         & 0938  &       & G1 Ib   &$-4.02 $& 6242. & 1.60 &$+0.17$& 1 &  \\
              & HD 170820         & 0939  &       & K0 III  &$-3.00 $& 4604. & 1.62 &$+0.17$& 1 &  \\
                  
M67           & M67 F 108         & 0919  &  058  &  GB     &$-0.26 $& 4255. & 1.84 &$-0.09$& 2 &L \\
                  
M71           & M71 1-09          & 0961  &  075  &  AGB    &$-0.42 $& 4672. & 1.72 &$-0.84$& 1 &L \\
              & M71 1-21          & 0958  &  076  &  GB     &$-0.69 $& 4364. & 1.53 &$-0.84$& 2 &L \\
              & M71 1-31          &  (*)  &  077  &         &$ 0.61 $& 5518. & 2.42 &$-0.84$& 1 &L \\
              & M71 1-34          & 0963  &  078  &  HB     &$ 0.74 $& 5075. & 2.43 &$-0.84$& 1 &L \\
              & M71 1-36          &  (*)  &  079  &         &$-0.92 $& 4682. & 1.42 &$-0.84$& 1 &L \\
              & M71 1-37          & 0959  &  080  &  GB     &$ 0.45 $& 4574. & 2.20 &$-0.84$& 1 &L \\
              & M71 1-39          & 0962  &  081  &  HB     &$ 0.81 $& 4976. & 2.45 &$-0.84$& 1 &L \\
              & M71 1-41          & 0960  &  082  &  HB     &$ 0.70 $& 5123. & 2.41 &$-0.84$& 1 &L \\
              & M71 1-53          & 0964  &  083  &  GB     &$-0.74 $& 4167. & 1.51 &$-0.84$& 1 &L \\
              & M71 1-59          &  (*)  &  084  &  GB     &$ 0.92 $& 4623. & 2.42 &$-0.84$& 1 &L \\
              & M71 1-63          & 0957  &  085  &  AGB    &$-0.18 $& 4689. & 1.87 &$-0.84$& 1 &L \\
              & M71 1-64          & 0956  &  086  &  GB     &$-0.61 $& 4275. & 1.59 &$-0.84$& 1 &L \\
              & M71 1-65          & 0955  &  087  &  GB     &$ 0.49 $& 4606. & 2.20 &$-0.84$& 1 &L \\
              & M71 1-66          & 0954  &  088  &  AGB    &$-0.70 $& 4465. & 1.55 &$-0.84$& 1 &L \\
              & M71 1-71          & 0951  &  089  &  GB     &$-0.17 $& 4404. & 1.84 &$-0.84$& 1 &L \\
              & M71 1-73          & 0949  &  090  &  GB     &$ 1.08 $& 4793. & 2.50 &$-0.84$& 1 &L \\
              & M71 1-75          & 0948  &  091  &         &$ 1.14 $& 4790. & 2.52 &$-0.84$& 2 &L \\ \hline
\end{tabular}
\end{center}
\end{table*}
\end{scriptsize}
\vfill\eject

\begin{scriptsize}
\begin{table*}                                                          
\begin{center}
\contcaption{}
\begin{tabular}{llcclrrrrcc}
\hline
Cluster name  & Star ID & MILES Num & CaT Num  & SpT     &$M_V^0$ & $T_{\rm eff}$ & $\log g$ & [Fe/H] & Ref($T_{\rm eff}$)&  Library \\
\hline
              & M71 1-77          & 0967  &       &         &$-1.00 $& 4014. & 1.32 &$-0.84$& 1 &  \\
              & M71 1-78          & 0968  &       &         &$-0.84 $& 4332. & 1.45 &$-0.84$& 1 &  \\
              & M71 1-87          & 0953  &  092  &         &$ 0.66 $& 5075. & 2.40 &$-0.84$& 1 &L \\
              & M71 1-95          & 0946  &  093  &  AGB    &$-0.41 $& 4639. & 1.73 &$-0.84$& 1 &L \\
              & M71 1-107         & 0947  &  094  &  AGB    &$-0.01 $& 4919. & 1.99 &$-0.84$& 1 &L \\
              & M71 1-109         & 0945  &  095  &  GB     &$ 1.16 $& 4723. & 2.55 &$-0.84$& 1 &L \\
              & M71 A2            & 0966  &  097  &  HB     &$ 0.82 $& 4840. & 2.35 &$-0.84$& 2 &L \\
              & M71 A4            &  (*)  &  098  &  AGB    &$-1.53 $& 4040. & 0.94 &$-0.84$& 2 &L \\
              & M71 A9            & 0944  &  101  &  GB     &$-0.78 $& 4151. & 1.45 &$-0.84$& 2 &L \\
              & M71 C             &  (*)  &  102  &  HB     &$ 0.78 $& 4892. & 2.36 &$-0.84$& 2 &L \\
              & M71 S             & 0969  &  104  &  GB     &$-0.78 $& 4247. & 1.45 &$-0.84$& 2 &L \\
              & M71 X             & 0970  &  105  &  HB     &$ 0.69 $& 5170. & 2.41 &$-0.84$& 2 &L \\
              & M71 I             & 0971  &       &         &$-1.33 $& 4275. & 1.10 &$-0.84$& 1 &  \\
              & M71 Y             &  (*)  &       &         &$ 0.16 $& 4649. & 2.04 &$-0.84$& 1 &  \\
              & M71 KC-147        & 0950  &  108  &         &$ 1.29 $& 4901. & 2.61 &$-0.84$& 1 &L \\
              & M71 KC-169        & 0965  &  109  &         &$ 0.59 $& 5014. & 2.36 &$-0.84$& 1 &L \\
              & M71 KC-263        & 0952  &  110  &         &$ 1.34 $& 4883. & 2.61 &$-0.84$& 1 &L \\
      
M79           & M79 131           &  (*)  &       &         &$-1.98 $& 4196. & 0.93 &$-1.37$& 2 &  \\
              & M79 153           & 0915  &       &         &$-2.15 $& 4130. & 0.79 &$-1.37$& 2 &  \\
              & M79 160           & 0916  &       &         &$-2.57 $& 3956. & 0.52 &$-1.37$& 2 &  \\
              & M79 223           & 0917  &       &         &$-2.40 $& 3977. & 0.64 &$-1.37$& 2 &  \\
              
M92           & M92 III 13        & 0937  &  114  &  GB     &$-2.61 $& 4178. & 0.77 &$-2.16$& 2 &L \\
              & M92 IV 114        & 0936  &  115  &  GB     &$-0.81 $& 4728. & 1.70 &$-2.16$& 2 &L \\
              & M92 IX 12         &  (*)  &  117  &  AGB    &$-0.52 $& 5677. & 1.87 &$-2.16$& 1 &L \\
              & M92 XII 8         &  (*)  &  118  &  GB     &$-1.88 $& 4477. & 1.18 &$-2.16$& 2 &L \\
              
NGC 288       & NGC 288 77        & 0897  &       &         &$-1.83 $& 4218. & 0.89 &$-1.07$& 2 &  \\
              & NGC 288 96        &  (*)  &       &         &$-2.21 $& 4023. & 0.65 &$-1.07$& 2 &  \\
              
NGC\,2420     & NGC 2420 140      & 0918  &       &         &$-0.82 $& 4397. & 1.73 &$-0.44$& 1 &  \\
              
NGC\,6791     & NGC 6791 R4       & 0940  &       &         &$ 0.28 $& 4072. & 1.71 &$+0.40$& 1 &  \\
              & NGC 6791 R5       & 0941  &       &         &$ 1.25 $& 4057. & 2.32 &$+0.40$& 1 &  \\
              & NGC 6791 R16      & 0942  &       & K4III   &$ 0.06 $& 4043. & 1.59 &$+0.40$& 1 &  \\
              & NGC 6791 R19      & 0943  &       &         &$ 0.46 $& 4086. & 1.86 &$+0.40$& 1 &  \\
              
NGC\,7789     & NGC 7789 329(491) & 0972  &       &         &$-0.07 $& 4527. & 2.14 &$-0.13$& 1 &  \\
              & NGC 7789 338(499) &  (*)  &  145  &         &$-0.52 $& 5120. & 1.90 &$-0.13$& 1 &L \\
              & NGC 7789 342(502) & 0974  &       &         &$ 0.10 $&10023. & 2.25 &$-0.13$& 1 &  \\
              & NGC 7789 353(509) & 0975  &       &         &$ 0.22 $& 4494. & 2.31 &$-0.13$& 1 &  \\
              & NGC 7789 415(550) & 0976  &  146  &  GB     &$-1.64 $& 4074. & 1.10 &$-0.13$& 1 &L \\
              & NGC 7789 461(583) & 0977  &       &         &$-1.02 $& 4243. & 1.56 &$-0.13$& 1 &  \\
              & NGC 7789 468(589) & 0973  &  147  &  GB     &$-1.19 $& 4273. & 1.39 &$-0.13$& 1 &L \\
              & NGC 7789 494(604) &  (*)  &       &         &$-1.69 $& 3930. & 1.04 &$-0.13$& 1 &  \\
              & NGC 7789 501(614) & 0978  &  149  &  GB     &$-1.13 $& 4115. & 1.48 &$-0.13$& 1 &L \\
              & NGC 7789 575(671) & 0979  &  150  &  GB     &$-0.24 $& 4544. & 2.06 &$-0.13$& 1 &L \\
              & NGC 7789 637(723) & 0980  &       &         &$ 0.09 $& 4561. & 2.24 &$-0.13$& 1 &  \\
              & NGC 7789 669(732) &  (*)  &       &         &$-0.87 $& 4214. & 1.65 &$-0.13$& 1 &L \\
              & NGC 7789 732(784) &  (*)  &       &         &$ 0.34 $& 4892. & 2.44 &$-0.13$& 1 &  \\
              & NGC 7789 765(804) & 0981  &       &         &$-0.73 $& 4578. & 1.76 &$-0.13$& 1 &  \\
              & NGC 7789 859(853) & 0982  &  153  &  GB     &$ 0.30 $& 4544. & 2.34 &$-0.13$& 1 &L \\
              & NGC 7789 866(864) &  (*)  &       &         &$ 0.51 $& 4853. & 2.51 &$-0.13$& 1 &  \\
              & NGC 7789 875(873) & 0983  &  154  &  HB     &$ 0.63 $& 4952. & 2.62 &$-0.13$& 1 &L \\
              & NGC 7789 897(881) & 0984  &  155  &  HB     &$ 0.56 $& 4912. & 2.58 &$-0.13$& 1 &L \\
              & NGC 7789 971(946) & 0985  &  156  &  GB     &$-1.31 $& 4020. & 1.32 &$-0.13$& 1 &L \\
\hline
\end{tabular}
\end{center}
\end{table*}
\end{scriptsize}

\end{document}